\documentclass[fleqn,usenatbib]{mnras}
\usepackage{amsmath}
\usepackage{xspace}
\usepackage{mathptmx}
\usepackage{txfonts}
\usepackage{xcolor}

\usepackage[T1]{fontenc}
\usepackage{ae,aecompl}

\usepackage{graphicx}	
\usepackage{amssymb}	
\usepackage{booktabs}
\usepackage{wasysym}
\usepackage{ulem}

\let\oldhref\href
\renewcommand{\href}[2]{\oldhref{#1}{\hbox{#2}}}

\definecolor{colorl1}{RGB}{0, 51, 153}
\definecolor{colorl2}{RGB}{153, 0, 0}
\definecolor{colorl3}{RGB}{179, 179, 0}
\definecolor{colorl4}{RGB}{51, 102, 0}

\definecolor{colorw1}{RGB}{51, 102, 255}
\definecolor{colorw2}{RGB}{255, 51, 0}
\definecolor{colorw3}{RGB}{255, 214, 51}
\definecolor{colorw4}{RGB}{51, 204, 51}
\definecolor{seagreen}{rgb}{0.190, 0.525, 0.361} 

\newcommand{\hMpc}{{\ifmmode{h^{-1}{\rm Mpc}}\else{$h^{-1}$Mpc}\fi}}
\newcommand{\Mpc}{{\ifmmode{{\rm Mpc}}\else{Mpc}\fi}}
\newcommand{\hkpc}{{\ifmmode{h^{-1}{\rm kpc}}\else{$h^{-1}$kpc}\fi}}
\newcommand{\kpc}{{\ifmmode{ {\rm kpc}}\else{{\rm kpc}}\fi}}
\newcommand{\kms}{{\ifmmode{ {\rm km\,s^{-1}}}\else{ ${\rm km\,s^{-1}}$}\fi}}
\newcommand{\hMsun}{{\ifmmode{h^{-1}{\rm {M_{\astrosun}}}}\else{$h^{-1}{\rm{M_{\astrosun}}}$}\fi}}
\newcommand{\Msun}{{\ifmmode{{\rm M}_{\astrosun}}\else{${\rm M}_{\astrosun}$}\fi}}
                        
\newcommand{\Mhalo}{{\ifmmode{M_{\rm halo}}\else{$M_{\rm halo}$}\fi}}
\newcommand{\Rvir}{{\ifmmode{R_{\rm vir}}\else{$R_{\rm vir}$}\fi}}
\newcommand{\Rtwohun}{{\ifmmode{R_{200}}\else{$R_{200}$}\fi}}
\newcommand{\Mvir}{{\ifmmode{M_{\rm vir}}\else{$M_{\rm vir}$}\fi}}
\newcommand{\Mtwohun}{{\ifmmode{M_{200}}\else{$M_{200}$}\fi}}
\newcommand{\Nvir}{{\ifmmode{N_{\rm vir}}\else{$N_{\rm vir}$}\fi}}
\newcommand{\Mstar}{{\ifmmode{M_{\rm star}}\else{$M_{\rm star}$}\fi}}
\newcommand{\Vrot}{{\ifmmode{V_{\rm rot}}\else{$V_{\rm rot}$}\fi}}
\newcommand{\ltsima}{$\; \buildrel < \over \sim \;$}
\newcommand{\gtsima}{$\; \buildrel > \over \sim \;$}
\newcommand{\lsim}{\lower.5ex\hbox{\ltsima}}
\newcommand{\gsim}{\lower.5ex\hbox{\gtsima}}

\def\lesssim{\mathrel{\hbox{\rlap{\hbox{\lower4pt\hbox{$\sim$}}}\hbox{$<$}}}}
\def\gtrsim{\mathrel{\hbox{\rlap{\hbox{\lower4pt\hbox{$\sim$}}}\hbox{$>$}}}}

\newcommand{\beq}{\begin{equation}}
\newcommand{\eeq}{\end{equation}}
\def\beqa{\begin{eqnarray}}
\def\eeqa{\end{eqnarray}}
\def\LCDM{\ensuremath{\Lambda}CDM}

\def\head{ \vbox to 0pt{\vss \hbox to 0pt{\hskip 440pt\rm
      LA-UR-10-07069\hss} \vskip 25pt}}

\def \kms {\ifmmode  \,\rm km\,s^{-1}\else $\,\rm km\,s^{-1}$\fi }
\def \kpc {\ifmmode  {\,\rm kpc}  \else ${\rm  kpc}$ \fi  }  
\def \hkpc {\ifmmode  {h^{-1}\rm kpc}  \else ${h^{-1}\rm kpc}$ \fi  }  
\def \hMpc {\ifmmode  {h^{-1}\rm Mpc}  \else ${h^{-1}\rm Mpc}$ \fi  }  
\def \Mpch {\ifmmode  {h^{-1}\rm Mpc}  \else ${h^{-1}\rm Mpc}$ \fi  }  
\def \Msun {\ifmmode {\rm M}_{\astrosun} \else ${\rm M}_{\astrosun}$ \fi} 
\def \hMsun {\ifmmode h^{-1}\,\rm M_{\astrosun} \else $h^{-1}\,\rm M_{\astrosun}$ \fi}
\def \Gyr {\ifmmode\, \rm Gyr \else $\,$Gyr \fi}

\def \LCDM {\ifmmode \Lambda{\rm CDM} \else $\Lambda{\rm CDM}$ \fi}
\def \sig8 {\ifmmode \sigma_8 \else $\sigma_8$ \fi} 
\def \OmegaM {\ifmmode \Omega_{\rm m} \else $\Omega_{\rm m}$ \fi} 
\def \Omegab {\ifmmode \Omega_{\rm b} \else $\Omega_{\rm b}$ \fi} 
\def \OmegaL {\ifmmode \Omega_{\rm \Lambda} \else $\Omega_{\rm \Lambda}$\fi} 
\def \Deltavir {\ifmmode \Delta_{\rm vir} \else $\Delta_{\rm vir}$ \fi}
\def \rhocrit {\ifmmode \rho_{\rm crit} \else $\rho_{\rm crit}$ \fi}
\def \rhou {\ifmmode \rho_{\rm u} \else $\rho_{\rm u}$ \fi}
\def \zc {\ifmmode z_{\rm c} \else $z_{\rm c}$ \fi}

\def\Mstar {\ensuremath {M_{*}(<r_{23.5})}~}
\def\r23_5 {\ensuremath {r_{23.5}}~}

\title[AGN effect on concentration and spin] {NIHAO XXVIII: Collateral effects of AGN on dark matter concentration and stellar kinematics}

\author[S. Waterval et al.]{Stefan Waterval$^{1,2}$\thanks{E-mail: sw4445@nyu.edu}, Sana Elgamal$^{1,2}$, Matteo Nori$^{1,2}$,  Mario Pasquato$^{1,2,3,4}$,
\newauthor{Andrea V. Macciò$^{1,2,5}$, Marvin Blank$^{1,2,6}$, Keri L. Dixon$^{1,2}$, Xi Kang$^{7,8}$, Tengiz Ibrayev}$^{1}$\\
$^{1}$New York University Abu Dhabi, PO Box 129188, Abu Dhabi, United Arab Emirates \\
$^2$Center for Astro, Particle and Planetary Physics (CAP$^3$), New York University Abu Dhabi\\
$^3$Physics and Astronomy Department Galileo Galilei, University of Padova, Vicolo dell’Osservatorio 3, I–35122, Padova\\
$^4$Département de Physique, Université de Montréal, Montreal, Quebec H3T 1J4, Canada\\
$^5$Max-Planck-Institut f\"ur Astronomie, K\"onigstuhl 17, 69117 Heidelberg, Germany\\
$^6$Institut f\"ur Theoretische Physik und Astrophysik, Christian-Albrechts-Universit\"at zu Kiel, Leibnizstr 15, D-24118 Kiel, Germany\\
$^7$Zhejiang University-Purple Mountain Observatory Joint Research Center for Astronomy, Zhejiang University, Hangzhou 310027, China\\
$^8$Purple Mountain Observatory, 10 Yuan Hua Road, Nanjing 210034, China\\
}

\date{Accepted XXX. Received YYY; in original form ZZZ}

\pubyear{2021}

\begin{document}

\label{firstpage}
\pagerange{\pageref{firstpage}--\pageref{lastpage}}
\maketitle
\begin{abstract}
Although active galactic nuclei (AGN) feedback is required in simulations of galaxies to regulate star formation, further downstream effects on the dark matter distribution of the halo and stellar kinematics of the central galaxy can be expected.
We combine simulations of galaxies with and without AGN physics from the Numerical Investigation of a Hundred Astrophysical Objects (NIHAO) to investigate the effect of AGN on the dark matter profile and central stellar rotation of the host galaxies. Specifically, we study how the concentration-halo mass ($c-M$) relation and the stellar spin parameter ($\lambda_R$) are affected by AGN feedback. 
We find that AGN physics is crucial to reduce the central density of simulated
massive ($\gtrsim 10^{12}\,\Msun$) galaxies and bring their concentration to agreement with results
from the \textit{Spitzer Photometry \& Accurate Rotation Curves} (SPARC) sample.
Similarly, AGN feedback has a key role in reproducing the dichotomy 
between slow and fast rotators as observed by the ATLAS$^{3\text{D}}$ survey.
Without star formation suppression due to AGN feedback, the number of fast rotators strongly exceeds the observational constraints. 
Our study shows that there are several collateral effects that support the 
importance of AGN feedback in galaxy formation, and these effects can be used to
constrain its implementation in numerical simulations.


\end{abstract}

\begin{keywords}
quasars: supermassive black holes, galaxies: formation, galaxies: evolution, methods: numerical, methods: statistical
\end{keywords}

\section{Introduction}\label{sec:introduction}




In recent years, astronomical observations have increasingly supported the idea that a large number of --~if not all~-- massive galaxies contain a supermassive black hole (BH) at their center, with masses ranging from $10^{5}$ to $10^{10}$ \Msun \citep[see, for review][and references therein]{cattaneo_2009_nature,harrison_2017}. Remarkably, observations and studies of supermassive BHs highlight surprising correlations between the mass of the central BHs and their host galaxies \citep[e.g][]{kormendy_2013_araa}. The presence of such correlations is not obvious as BHs and galaxies differ by several orders of magnitude in physical size scales, thus suggesting that the evolution of the two elements may be closely correlated due to a significant and mutual influence \citep[but see also][for a possible different interpretation]{JahnkeMaccio2011}.

Accretion onto a BH gives rise to multiple observable phenomena, including electromagnetic radiation, relativistic jets, and less-collimated, non-relativistic outflows \citep[][]{krolik_1999}. By emitting large amounts of energy and momentum, \textit{active galactic nuclei} (AGN) can have a significant effect on the formation of stars in the galaxy: the energy and momentum released can couple with the gas in and around the galaxy through various physical mechanisms. Specifically, AGNs can heat up the gas around them, thereby providing a form of thermal feedback, while kinetic feedback is provided by driving winds that eject gas \citep[e.g.][]{King2003}.

In fact, the liberation of energy in the neighbouring environment by AGNs --~commonly referred to as AGN feedback~-- is needed in most galaxy formation simulations to reduce star formation and recover various key observables of massive galaxies \citep[e.g.][]{valageas_1999_aap,croton_2006_mnras,somerville_2008_mnras,vogelsperger_2014_mnras,crain_2015_mnras,costa_2018_mnras,nihao22,zinger_2020_mnras}. The energy that is transferred from the AGN to its environment effectively heats up the surrounding gas and counters the cooling process required for the formation of new stars. As a result, the overall age distribution of stars in galaxies hosting AGN skews towards a more old and therefore red population \citep[e.g.][]{thomas_2005_apj}.


Some evidence of the quenching of star formation caused by AGN feedback derives from observations of a small number of distant luminous AGN at redshifts $z \sim 1-3$, where ionized outflows were found to be spatially anti-correlated with the location of narrow H$\alpha$ emission regions, which constitute one of the star-formation tracers \citep[][]{cano_diaz_2012_aap,carniani_2016_aap}. Statistical studies on large samples of AGN-hosting galaxies have also systematically linked AGN feedback and star formation rate (SFR), with radio-loud sources being  consistently found to have low SFR \citep[][]{hardcastle_2013_mnras,ellison_2016_mnras,leslie_2016_mnras,ellison_2016_mnras,comerford_2020_apj}. AGN feedback has also been linked to the quenching of star formation in the most luminous AGNs \citep[][]{page_2012_nat}. It is worth noting that some tension still remains, with some observational studies making opposite claims and revealing evidence of star formation enhancement due to AGN activity \citep[][]{elbaz_2009_aap,lutz_2010_apj,santini_2012_aap,juneau_2013_apj,bernhard_2016_mnras,dahmer-hahn_2022_mnras} or suppression and enhancement working synchronously in the same galaxy \citep[][]{zinn_2013_apj,karouzos_2014_apj,cresci_2015_apj,shin_2019}.

While AGN feedback is mainly needed and used as a means to stop star formation in simulations of galaxy formation and evolution, 
it also has some collateral effects on the dynamics and distribution of collisionless components of a galaxy: stars and dark matter (DM). \citet{martizzi_2012_mnras} compared numerical simulations of galaxy clusters with and without AGN feedback and obtained flat density cores in DM and stellar profiles when AGN feedback was included, in contrast to predictions of DM-only  simulations where a so-called \textit{cusp} is
expected \citep[e.g.][]{dubinski_1991_apj,navarro_1997_apj}.
 More recently, \citet{nihao23} have also shown that AGN-induced gas outflows can act against the natural DM contraction caused by the presence of a large stellar component in the centre of massive haloes \citep[][]{blumenthal_1986_apj,Gnedin2004,Abadi2010,schaller_2015_mnras}.
A common way to compare DM distribution in simulated and real objects is via the concentration-mass ($c-M$) relation \citep[e.g][]{Bullock2000,Maccio2007}, since
the concentration parameter, $c$, can be easily computed from galaxy rotation curves.

Another impact of AGNs on galactic properties, complementary to the effect on star formation, is the repercussion on the  kinematic properties of the stellar body.
Elliptical galaxies have been found to have more complex
kinematics than initially thought. 
\citet{sauron_ix} used 2D stellar kinematics of elliptical and lenticular galaxies using the SAURON integral-field spectrograph \citep[][]{sauron_i} and found significant coherent rotation in 
the inner part of these galaxies.
They hence proposed a new classification scheme for early-type galaxies: \textit{slow rotators} (SRs) and \textit{fast rotators} (FRs), according to the amount of large-scale rotation present.
Their results were confirmed and extended by the ATLAS$^{\text{3D}}$ \citep[][]{atlas3d_i} survey that analyzed 260 early-type galaxies and classified
86~per~cent of them as FRs based on the measurements of their spin parameter corrected for ellipticity \citep[][]{atlas3d_iii}.
Recently, \citet{frigo_2019_mnras} used 20 cosmological simulations of massive galaxies to study the effects of  AGN on galaxy rotation, finding that indeed
AGN feedback  enhances the production of SRs due to the reduced in-situ star formation in the  galaxies.

In this manuscript, we intend to study the aforementioned collateral impacts of AGNs using simulated galaxies from the \textit{Numerical Investigation of a Hundred Astrophysical Objects} \citep[NIHAO;][]{nihao1}. It contains a large statistical sample ($\sim 150$) of high-resolution cosmological zoom-in simulations, ranging from dwarf to elliptical galaxies. NIHAO simulations include metal cooling, star formation, chemical enrichment, and feedback from massive stars, supernovae, and AGNs. Each galaxy is resolved with $\sim 10^6$ particles, resolving mass profiles down to 1~per~cent of the virial radius.

NIHAO simulations successfully recover various galaxy properties such as the stellar-to-halo mass relation (SHMR) \citep[][]{nihao1}, the disk gas mass and disk size relation \citep[][]{nihao10}, the Tully-Fisher relation \citep[][]{nihao12}, the diversity of dwarf galaxy rotation curves \citep[][]{nihao14}, and the satellite mass function of the Milky Way and M31 are well recovered by the NIHAO simulations \citep{nihao15}, as well as the star formation main sequence of star-forming galaxies \citep[][]{nihao_26}. This makes the NIHAO suite the optimal tool to test the $c-M$ relation and compare it with observations, as well as investigate the effect of AGN feedback on $c$ and galaxy stellar kinematics.

In the first part of this paper, we investigate the effect of AGN on DM concentration in simulated galaxies from NIHAO. We compute the $c-M$ relation of 143 NIHAO galaxies and compare it with the $c-M$ relation inferred from the rotation curves of 175 late-type galaxies from the \textit{Spitzer Photometry \& Accurate Rotation Curves} \citep[SPARC;][hereafter L20]{sparc3}. We test five out of the seven profiles used in L20 for galaxy masses covering five decades in range and use the halo masses obtained from Markov Chain Monte Carlo (MCMC) fitting to compute the $c-M$ relation of the NIHAO sample.

In the second part of this paper, we perform a similar analysis
as in \citet[][]{frigo_2019_mnras}. Starting from the NIHAO database, we choose a subset of 40 massive elliptical galaxies and probe the abundances of SRs and FRs. We confirm that the characteristics of this distribution are in agreement with observations from ATLAS$^{\text{3D}}$ \citep[][]{atlas3d_i}. Furthermore, we trace the stellar kinematics properties of a subset of five galaxies with and without AGN physics, which allows us to infer the effect of AGN feedback on the evolution of the kinematics of the stellar component.

This paper is organized as follows: in Section~\ref{sec:methods}, we first review the theoretical background used in this work before describing the NIHAO simulation suite as well as the observed SPARC sample to which we compare our simulations. We end this Section by introducing the methods to extract the necessary quantities from the simulations and how the analysis on the dark and stellar parts are performed. Our results are then presented in Section~\ref{sec:results}, while Section~\ref{sec:discussion} is devoted to summarizing and discussing our findings.

\section{Methods}\label{sec:methods}

This section is dedicated to the methods we used to perform our analysis. We begin by providing some theoretical background motivating our calculations before introducing the data used in this work. We then continue by explaining how the intrinsic virial radius is extracted from simulations, as well as the resulting intrinsic virial mass. Furthermore, we detail the MCMC procedure used to fit the different DM density models to the data and describe how the `observed' virial radius and virial mass, as well as their respective errors are calculated. We then outline the statistical method we choose to compare with observations from SPARC before finally explaining how we extract the spin parameter and ellipticity of the subset of early-type galaxies from the NIHAO catalogue. All of the relevant quantities in this section are extracted from simulations using the \textsc{\small PYNBODY} \citep[][]{pynbody} post-processing tool.

\subsection{Theoretical background}\label{sec:theory}


In this subsection, we briefly review the most common equations that are used in the literature to represent DM radial profiles of collapsed structures as well as the theoretical definition of DM concentration. We also revise the definition of the dynamical properties relevant for the dynamical and morphological analysis of the stellar component in this work: $\lambda_{R}$ and ellipticity.

\subsubsection{DM profiles}\label{subsubsec:dmprofiles}


Following L20, we use a set of five different functional forms to fit the DM density profiles extracted from the NIHAO simulations, namely:
Navarro-Frenk-White \citep[][hereafter NFW]{navarro_1996_apj}, Einasto \citep[][]{einasto_1965}, pseudo-isothermal (pISO), Burkert \citep[][]{burkert_1995_apj} and Lucky13 \citep[][]{sparc3}.
For each of these profiles, we define the concentration parameter $c$ as the ratio between the virial radius of the halo (computed at a density
of 200 times the critical density of the Universe) and the scale radius ($r_s$) that describes the corresponding analytic profile.
The $c-M$ relation has been extensively studied for $N$-body simulations \citep[e.g.][]{navarro_1997_apj,Bullock2001,Maccio2007,Neto2007,Prada2012,Dutton2014}, and several fitting formulae have been provided. In order to be consistent with L20, we will use here the predictions suggested by \citet[][]{Maccio2008}.

\paragraph{NFW}

The NFW profile was one of the first attempts to have a universal description of DM-halo density profiles from $N$-body (gravity only) simulations. This density profile reads:
\begin{equation}\label{eq:nfw_profile}
\rho_{\text{NFW}} = \frac{\rho_{\text{s}}}{x\left(1+x\right)^2},
\end{equation}
and its enclosed mass:
\begin{equation}\label{eq:nfw_mass}
M_{\text{NFW}} = 4\,\pi\,\rho_{\text{s}}\,r_{\text{s}}^3\,\left[\ln(1+x)-\frac{x}{1+x}\right],
\end{equation}
where we use the dimensionless parameter $x=r/r_{\text{s}}$. $r_{\text{s}}$ and $\rho_{\text{s}}$ represent the radius where the logarithmic slope changes from -1 to -3, and the characteristic density of the halo at $r_{\text{s}}$, respectively. Note that the NFW profile has an inner slope of $-1$ and an outer slope of $-3$.

\paragraph{Einasto}

\citet{navarro_2004_mnras} proposed a refined version of the original NFW profiles, based on the Einasto model \citep{einasto_1965}. This profile introduces an additional shape parameter ($\alpha$) leading to a versatile profile that can adapt to different shapes of central densities:
\begin{equation}\label{eq:ein_profile}
\rho_{\text{Einasto}} = \rho_{\text{s}} \ e^{-\frac{2}{\alpha \left[ x^{\alpha}-1 \right]}},
\end{equation}
and the enclosed mass is given by
\begin{equation}\label{eq:ein_mass}
M_{\text{Einasto}} = 4\,\pi\,\rho_{\text{s}}\,r_{\text{s}}^3\,\left(\frac{2}{\alpha}\right)^{-\frac{3}{\alpha}}\frac{e^{2/\alpha}}{\alpha}\,\Gamma\left(\frac{3}{\alpha},\frac{2}{\alpha}x^{\alpha}\right),
\end{equation}
with $\Gamma$ representing the incomplete Gamma function.

\paragraph{pISO and Burkert}

The pISO model has proven to accurately reproduce the rotation curves of dwarf galaxes \citep{adams_2014_apj,oh_2015_aj}. This model assumes a constant density core of radius $r_s$ and density $\rho_s$ and is expressed as
\begin{equation}\label{eq:piso_profile}
    \rho_{\text{pISO}} = \frac{\rho_{\text{s}}}{1+x^2},
\end{equation}
and its enclosed mass is
\begin{equation}\label{eq:piso_mass}
   M_{\text{pISO}} = 4\,\pi\,\rho_{\text{s}}\,r_{\text{s}}^3\,[x-\arctan(x)].
\end{equation}
One caveat with pISO is that the enclosed mass quickly diverges at large radii. \citet{burkert_1995_apj} proposed a profile that diverges more slowly than pISO
\begin{equation}\label{eq:burkert_profile}
    \rho_{\text{Burkert}} = \frac{\rho_{\text{s}}}{\left(1+x\right) \left(1+x^2\right)}
\end{equation}
with enclosed mass
\begin{equation}\label{eq:burkert_mass}
M_{\text{Burkert}} = 2\,\pi\,\rho_{\text{s}}\,r_{\text{s}}^3\,\left[\frac{1}{2}\ln(1+x^2)+\ln(1+x)-\arctan(x)\right].
\end{equation}

\paragraph{Lucky13}

Finally, L20 suggest a new model based on the ($\alpha$, $\beta$, $\gamma$) models \citep{hernquist_1990,zhao_1996_mnras}, where $\alpha$ is set to 1, $\beta$ to 3, and $\gamma$ to 0. Lucky13 hence reproduces a finite core near the center and a $-3$ NFW slope at large radii. The profile is given by

\begin{equation}\label{eq:lucky13_profile}
    \rho_{\text{Lucky13}} = \frac{\rho_{\text{s}}}{\left(1+x\right)^3},
\end{equation}
and the corresponding enclosed mass
\begin{equation}\label{eq:lucky13_mass}
    M_{\text{Lucky13}} = 4\,\pi\,\rho_{\text{s}}\,r_{\text{s}}^3\,\left[\ln(1+x)+\frac{2}{1+x}-\frac{1}{2(1+x)^2}-\frac{3}{2}\right].
\end{equation}



\subsection{Data}\label{sec:data}

The NIHAO sample of 143 galaxies used is divided into subgroups according to whether or not AGN physics was included in the simulation. The galaxies strictly without AGN are referred to as `NoAGN' and constitute 11 galaxies, while simulations including BH growth, accretion, and feedback make up 41 elements in the sample and are labeled `AGN'. In addition to these, 91 galaxies have been run both with and without BH and present therefore a suitable sample to quantify the effect of AGN on the DM distribution and stellar kinematics of galaxies and their host halo. In the following two subsections, we present the NIHAO simulation suite as well as the sources of the observational data used for comparison with the simulations.

\subsubsection{NIHAO}\label{subsec:nihao}

The NIHAO project was initially aimed at producing a sample of $\sim\!100$ high-resolution galaxies covering three orders of magnitude in halo mass ($\sim\!10^{9}-10^{12}\,\Msun$), using cosmological zoom-in hydrodynamical simulations to study the formation and evolution of galaxies in a full cosmological framework \citep[][]{nihao1}. This suite uses the \textsc{\small GASOLINE2} \citep{wadsley_2017_mnras} code, and each halo contains approximately $\Nvir \sim 10^{6}$ particles within the virial radius, \Rvir. The adopted cosmological framework is a flat $\Lambda$CDM cosmology with parameters from \citet{planck_2013}. The Hubble parameter $H_0 = 67.1 \kms\Mpc^{-1}$ and the matter, dark energy, radiation, and baryon densities are $\{\Omega_{\text{m}},\Omega_{\Lambda},\Omega_{\text{r}},\Omega_{\text{b}}\} = \{0.3175,0.6824,0.00008,0.0490\}$. The power spectrum normalisation and slope are $\sigma_8 = 0.8344$ and $n = 0.9624$, respectively. Haloes in NIHAO simulations are identified using the \textsc{\small Amiga Halo Finder} \citep[\textsc{\small AHF};][]{gill_2004_mnras,knollmann_2009_apj}.

Initial conditions are produced with a modification of the \textsc{\small GRAFIC2} code \citep[][]{Bertschinger2001}. Details about the modifications can be found in \citet{Penzo2014}. The refinement level is set in such a way to keep the ratio between the DM softening $\epsilon_{\text{DM}}$ and \Rvir to approximately $0.003$. This constant relative resolution resolves the DM mass profile down to 1~per~cent of \Rvir.

Star formation follows the Kennicutt-Schmidt law \citep[][]{schmidt_1959_apj,kennicutt_1998_apj}, where gas particles exceeding a certain temperature and density threshold turn into star particles. The respective thresholds are $T < 15000$ K for the temperature and $n_{\text{th}} > 10.3$ cm$^{-3}$ for the density.

Different types of energy feedback are included in NIHAO. Gas is allowed to cool via negative feedback from Compton cooling and photoionization from the uniform ultraviolet background is implemented according to \citet{Haardt2012}. Stellar feedback is governed by two different phenomena: massive stars provide ionizing feedback before turning into supernovae, denoted as `early stellar feedback' \citep[][]{Stinson2013}, while supernova feedback energy is injected through blast-wave shocks \citep[][]{Stinson2006}.

More recently, BH seed, accretion, and feedback were incorporated in the NIHAO suite to study more massive elliptical galaxies, whose central AGNs are thought to play a crucial role in the quenching of star formation observed in elliptical galaxies \citep[see, e.g.][]{mcnamara_2007_araa}. In the following, we provide a summary of how AGN evolve and feed energy back to the system in NIHAO; a complete description can be found in \citet{nihao22}.

If a halo exceeds a threshold mass of $5 \times 10^{10}\,\Msun$, a BH seed particle of initial mass $M_{\text{BH,s}} = 1 \times 10^{5}$ \Msun is converted from the gas particle with the lowest gravitational potential. BH accretion and feedback are modeled as introduced by \citet{springel_2005_mnras}. Specifically, NIHAO uses the standard Bondi accretion model \citep[][]{bondi_1952_mnras} with the boost parameter set to 70. The accretion is capped to the Eddington rate, i.e. at each time-step $\Delta t$, the Bondi accretion rate ($\dot{M}_{\text{Bondi}}$) and the Eddington accretion rate ($\dot{M}_{\text{Edd}}$) are calculated and the accretion rate of the BH is taken as $\min\{\dot{M}_{\text{Bondi}},\dot{M}_{\text{Edd}}\}$.

During each time-step, the BH accretes the mass $\dot{M}_{\text{BH}}\Delta t$ from the most gravitationally bound gas particle to it. Once a gas particle reaches a mass below 20~per~cent of its initial mass, the particle is removed, and its mass and momentum are distributed among the neighbouring gas particles. There is no maximum distance at which a gas particle has to be in order to be accreted; but since it is the most bound particle, it is usually also the closest to the BH. The BH luminosity ($L_{\text{BH}}$) is computed from the accretion rate assuming a radiative efficiency of 10 per~cent \citep[][]{shakura_1973}, and a fraction of five per~cent of $L_{\text{BH}}$ is then assumed to be available as thermal energy for the surrounding environment, which is then distributed to the nearest 50 gas particles.

\subsubsection{SPARC}\label{subsec:sparc}

SPARC is an observational database comprising of 175 late-type nearby galaxies (S0 to Irr) with near-infrared surface photometry at 3.6 $\mu$m and extended HI rotation curves \citep[][]{sparc1}. The wide range in luminosity, surface brightness, and rotation velocity offers a good sample of disk galaxies in the nearby Universe and is thus appropriate for a comparative study with numerical simulations from NIHAO. Most of the observations come from \textit{The Spitzer Survey of Stellar Structure in Galaxies} \citep{sheth_2010}.

L20 performed MCMC rotation curve fits of the 175 SPARC galaxies using seven DM halo profiles: NFW, Einasto, pISO, Burkert, Di Cintio \citep[DC14;][]{dc_2014b_mnras}, coreNFW, and a new profile that the authors refer to as Lucky13. The fits are performed by summing each component of the observed rotation velocities $V_{\text{obs}}$ (DM, disk, bulge, and gas):

\begin{equation}
    V_{\text{tot}}^2 = V_{\text{DM}}^2 + \Upsilon_{\text{disk}} V_{\text{disk}}^2 + \Upsilon_{\text{bul}} V_{\text{bul}}^2 + V_{\text{gas}}^2.
\end{equation}

The DM profiles have two~(three) free parameters: $V_{200}$ and concentration $c_{200}$~(Einasto has an additional shape prameter $\alpha$), and the baryonic contributions have three free parameters: stellar mass-to-light ratio $\Upsilon$, galaxy distance $D$, and disk inclination $i$. These parameters therefore lead to a five~(six for Einasto) dimensional parameter space.

The authors impose both flat priors (for all halo profiles) and $\Lambda$CDM priors (for NFW, EInasto, DC14, coreNFW, and Lucky13) in their analysis. The $\Lambda$CDM priors comprise the SHMR from abundance matching \citep{moster_2013_mnras} and the $c-M$ relation from \citep{maccio_2008_mnras}.

L20 kindly provided their final processed data, we thus compare our results with L20 using their calculated halo masses and concentrations (see Fig. \ref{fig:m200c200}). For the NFW, Einasto, and Lucky13 profiles, we compare NIHAO against SPARC with $\Lambda$CDM priors, for consistency with our simulation suit.





\subsection{DM distribution}\label{subsec:dmdistribution}

\subsubsection{Virial radius and virial mass from simulations}\label{subsubsec:virialrad_virialmass}

In place of the proper cosmological definition of the virial radius, we hereafter use $R_{200}$, defined as the radius within which the halo contains an average density 200 times the critical density $\rho_{\text{c}}$ today, i.e. at $z=0$. The advantage of this definition is that it is independent of cosmological parameters. Note that, in the following, we discursively refer to this radius and related quantities as \textit{virial} quantities, but we keep the subscript $200$ for consistency. For a review on different ways to define the mass of a halo, see e.g. \cite{white_2001_aa}.

We start by calculating the intrinsic virial radius ($R_{200}^{\text{true}}$) of each halo (i.e. extracted directly from the simulation) and use it to compute the DM density profile from 1~per~cent to 20~per~cent of $R_{200}^{\text{true}}$. Observations only have access to the luminous fraction of the halo (the galaxy itself), and we therefore want to emulate the lack of directly available information exceeding a certain distance from the centre of the galaxy by restricting the range of our profile. The density is computed in 50 bins equally spaced in logarithmic scale. The error in each bin is estimated as the Poisson noise related to the finite number of particles in each bin. From the virial radius extracted from \textsc{\small PYNBODY}, we also calculate the intrinsic virial mass ($M_{200}^{\text{true}}$) which is the mass enclosed within $R_{200}^{\text{true}}$ as follows:

\begin{equation}
    M_{200}^{\text{true}} = \frac{4}{3}\,\pi\,(R_{200}^{\text{true}})^3\,200\,\rho_{\text{c}}.
\end{equation}

\subsubsection{MCMC fitting}\label{subsubsec:mcmc}

We use an MCMC method to fit the five different models to the density profiles of the NIHAO simulations. In particular, we use the MCMC python package \textsc{\small emcee} \citep{emcee}.

In MCMC analyses, the exploration of the multidimensional space of parameters is guided by the choice of the probability functional associated to every point, called \textit{likelihood}. In this regard, the region of parameter space to explore can be either effectively restricted with boundaries or differentially prioritized by defining an a priori probability function, i.e. the \textit{prior}. While L20 impose a set of $\Lambda$CDM priors to fits, we do not, since the $\Lambda$CDM cosmology is already embedded by construction in the data extracted from the NIHAO simulations. We set the boundaries for the free parameters $r_{\text{s}}$ and $\rho_{\text{s}}$ to $[0,R_{200}]$~\kpc and $[0,2\,\rho_{\text{max}}]\ \Msun\kpc^{-3}$, respectively, where $R_{200}$ is the virial radius obtained directly from the simulation and $\rho_{\text{max}}$ is the maximum bin density of the density profile of each galaxy. We choose the same likelihood function as L20, i.e. $\exp{(-\frac{1}{2}\chi^2})$ with $\chi^2$ having the standard definition
\begin{equation}
    \chi^2 = \sum_{N_\text{bins}} \frac{(\rho_i-\rho_i^{\text{fit}})^2}{(\delta \rho_i)^2},
\end{equation}
where $\rho_i$ is the density computed at bin $i$, $\rho_i^{\text{fit}}$ is the fitted model, and $\delta \rho_i$ is the error on $\rho_i$ due to Poisson noise. The MCMC chains are initialized with 200 random walkers and then run for a burn-in period of 300 iterations before the full run of 1000 iterations. We also check that the acceptance fractions approximately lie between 0.1 and 0.7.

\subsubsection{Virial radius and virial mass from DM profiles}\label{subsubsec:error}
The virial quantities needed to compute the $c-M$ relation are calculated using the parameters $r_{\text{s}}$ and $\rho_{\text{s}}$ resulting from the MCMC fit. Requiring that $\rho = \frac{M_{\text{enc}}}{4/3\,\pi\,r^3}$ be equal to $200\,\rho_{\text{crit}}$, where $M_{\text{enc}}$ represents the enclosed mass of a given profile (see \ref{subsubsec:dmprofiles}), allows for the determination of $R_{200}$ through simple binary search. To estimate the errors in our $c-M$ relation, we extract the $16^{\text{th}}$, $50^{\text{th}}$, and $84^{\text{th}}$ percentiles on $r_s$ and $\rho_s$ from the MCMC fit. The $50^{\text{th}}$ percentile values are used to calculate $R_{200}$, $M_{200}$, and $c_{200}$, while the spread of the upper and lower errors in both $M_{200}$ and $c_{200}$ make use of the $16^{\text{th}}$ and $84^{\text{th}}$ results.

\subsubsection{Statistical comparison with SPARC}\label{subsubsec:comparison_sparc}
The quantitative comparison between the $c-M$ relation found in this work and the one found in L20 is done by computing the likelihood $\mathcal{L}$ of the NIHAO data under the distribution $f_{\mathrm{SPARC}}(c, M)$ of SPARC data: 
\begin{equation}\label{eq:likelih}
    \mathcal{L} = \prod_i f_{\mathrm{SPARC}}(c_i, M_i),
\end{equation}
where  each $i$ represents one NIHAO data point in the $c-M$ relation (see Fig. \ref{fig:m200c200}).
The distribution $f_{\mathrm{SPARC}}(c, M)$ was estimated via Kernel Density Estimation (KDE), as implemented by the \textsc{\small KernelDensity} method from the \textsc{\small python} package scikit-learn \citep{sklearn}. Given $N$ data points $\mathbf{X}_i \in \mathbb{R}^N$, the KDE method estimates their distribution as
\begin{equation}
f\left(\mathbf{X}\right) = \frac{1}{N} \sum^{N}_{i = 1} K\left(\mathbf{b}, \mathbf{X} -\mathbf{X_i}\right),
\end{equation}
where $\mathbf{X} \in \mathbb{R}^N$ and $K: \mathbb{R}^N_{+} \times \mathbb{R}^N \to [0, 1]$ is called a kernel function of bandwidth $\mathbf{b}$. In the above equation, $K$ is normalized to $1$. In practice, $K$ is often chosen to be a $N$-dimensional Gaussian with $\mathbf{b}$ proportional to its standard deviation in each coordinate.
Bandwidth is a key parameter for KDE: a small value of the bandwidth results in an estimated $f$ that displays high variance, being overly sensitive to the idiosyncratic patterns of the data points $\mathbf{X}_i$; while a large one will produce a biased estimate of $f$, missing out on subtle details of the distribution of the data at small scales. 
In our analysis, the bandwidth is a one-dimensional parameter estimated from the errors in the SPARC data. Since the uncertainties in $\log(M_{200})$ and $\log(c_{200})$ are of the same order of magnitude, we chose to compute the average of all errors (restricted within the analysis domain of choice, see below) in both dimensions to determine the bandwidth to use for each profile.

To reduce the effect of outliers in the SPARC data, we start by removing all points lying outside the $2.5^{\text{th}}$ and $97.5^{\text{th}}$ percentiles in each dimension. We then use the \textsc{\small ConvexHull} method from the \textsc{\small SciPy} package \citep[][]{scipy} to select the outermost points of the remaining SPARC galaxies. These points are used to fit an ellipse with the \textsc{\small EllipseModel} class from the \textsc{\small scikit-image} package \citep[][]{skimage}, and the ellipse obtained in the $c-M$ relation of each profile constitutes the domain on which everything that follows is calculated.

After obtaining an estimate of $f_{\mathrm{SPARC}}$, we calculate $\mathcal{L}_{\mathrm{NIHAO}}$ for the NIHAO data following equation~(\ref{eq:likelih}) (for ease of computation we actually calculate $\log{\mathcal{L}}$), which we then treat as a statistic under a bootstrap approach.
For each profile, we extract a random set of $N$ (corresponding to the number of NIHAO galaxies within the domain delimited by the ellipse) uniformly distributed points lying within the ellipse in the $c-M$ plane and calculate $\log{\mathcal{L}_k}$ for each of them. This process is repeated for $10^5$ iterations. We then compared the score of the NIHAO data $\mathcal{L}_{\mathrm{NIHAO}}$ to these and calculate the fraction of the $10^5$ scores that fall above it. This bootstrap process returns a $p$-value for the null hypothesis that the NIHAO data is extracted at random.


\subsection{Stellar kinematics}\label{subsec:stellarkinematics}

\subsubsection{2D stellar kinematic maps}\label{subsubsec:2dstellarkinematicmaps}

From the AGN sample, we extract elliptical galaxies. Given our small sample, we visually inspect the stellar component of our galaxies at $z=0$ in edge-on view and exclude all simulations presenting disc-like features. We are left with 45 massive elliptical galaxies from the initial AGN sample. For each one, we spatially centre the stellar component using a shrinking sphere method following \citet[][]{power_2003_mnras}. In this algorithm, the center of mass of all the star particles is computed iteratively within a sphere of some large enough initial radius. At each iteration, the center is set equal to the last barycentre, and the radius of the sphere is reduced by 2.5~per~cent. The iteration stops when the sphere contains a specified number of star particles, here set to 100.

We then divide the central region of our simulations in a 60-by-60 pixels grid and construct 2D stellar line-of-sight velocity and velocity dispersion maps along a given direction by averaging each quantity contained in every pixel. Fig. \ref{fig:velmaps} show these 2D maps for the `edge-on' projection (i.e., perpendicular to the angular momentum vector of the stars) for two of our galaxies. 

\subsubsection{Spin parameter and ellipticity}\label{subsubsec:spin_ellipticity}

Traditionally, the amount of rotation of the stellar component has been quantified by the observed rotational velocity over the velocity dispersion, $v/\sigma$. \citet{sauron_ix} noted however that $v/\sigma$ failed to discriminate between certain galaxies exhibiting the same $v/\sigma$ and ellipticity $\epsilon$ but having different velocity fields. The authors show an example of two galaxies (NGC 3379 and NGC 5813), which both exhibit similar $v/\sigma$ and ellipticity. Their respective stellar velocity fields are, however, distinct with the first galaxy exhibiting a regular and large-scale rotation pattern, while the second one displays a central kinematically decoupled component. This component is amplified by the luminosity weighting in the computation of $v/\sigma$. To overcome this degeneracy, the authors introduced a new quantity, $\lambda_R$, providing a measurement of the stellar angular momentum of a galaxy from a 2D field.

To characterize the amount of rotation in the stellar body of massive galaxies, we calculate $\lambda_R$ as follows:
\begin{equation}\label{eq:lambdar}
    \lambda_R \equiv \frac{\langle R |v|\rangle}{\langle R \sqrt{v^2 + \sigma^2} \rangle} = \frac{\sum_i\,m_i\,R_i\,|v_i|}{\sum_i\,m_i\,R_i \sqrt{v_i^2 + \sigma_i^2}},
\end{equation}
where $R$ is the projected distance to the galactic center and the brackets $\langle ~\rangle$ denote a luminosity-weighted sky average. For the last equality, we assume a constant mass-to-light ratio and are therefore able to convert the flux-weighted average to the stellar mass-weighted average. The index $i$ runs over all the pixels, and $m_i$, $R_i$, $v_i$, and $\sigma_i$ denote the stellar mass, the projected distance from the center of the map, the average stellar line-of-sight velocity, and the average stellar line-of-sight velocity dispersion for a given pixel $i$, respectively.

We limit the sum to the pixels that lie inside one stellar projected half-mass isophote for better comparison with observations. The apertures are constructed following the method employed by \citet{penoyre_2017_mnras}. Starting from the most massive pixel at the center of the map, the next most massive pixel adjacent to it is added, and this process is repeated until the stellar mass of the included pixels exceeds half the total projected stellar mass enclosed within 10~per~cent of $R_{200}^{\text{true}}$.

To measure the ellipticity for each galaxy in our sample, we first construct the best-fitting ellipse using the center of the pixels at the edge of the constructed aperture, and infer the (photometric) major and minor axis. The best-fitting ellipse is determined following a method\footnote{\url{https://github.com/ndvanforeest/fit_ellipse}} based on \citet[][]{fitzgibbon_1996}. This procedure allows us to compute the ellipticity as follows:

\begin{equation}
    \epsilon \equiv 1 - \sqrt{\frac{\langle y^{2} \rangle}{\langle x^{2} \rangle}} = 1 - \sqrt{\frac{\sum_i\,m_i\,y_i^{2}}{\sum_i\,m_i\,x_i^{2}}},
\end{equation}
where in the latter equality we once again convert from a flux-weighted average to a stellar mass-weighted average. $y_i$ and $x_i$ are the projected distances from a given pixel $i$ to the major and minor axis, respectively.

Originally, a cutoff value of $\lambda_R = 0.1$ was proposed to discriminate early-type galaxies between SRs and FRs by \citet{sauron_ix}. This cutoff was further improved to depend on the apparent ellipticity $\epsilon$ \citep[][]{atlas3d_iii}, which ensures that the kinematic classification of early-type galaxies is independent of the viewing angle. This refined criterion, which we adopt in this work to classify FRs and SRs, is given by: 
\begin{equation}\label{eq:lambda_crit}
    \lambda_R = 0.31 \sqrt{\epsilon}.
\end{equation}

Hereafter, we denote our measured spin parameters and ellipticities by $\lambda_{R_{1/2}}$ and $\epsilon_{1/2}$, respectively, to emphasize that both quantities are computed only within the central region enclosing roughly half the total projected stellar mass. For each galaxy in our sample, the spin parameters and ellipticities are computed in one random projection (see Fig. \ref{fig:lambda_epsilon}). For a few selected galaxies, both quantities are computed in edge-on projection to study the evolution of the spin parameter as a function of radius and redshift (see Figs. \ref{fig:lambda_profile_radius} and \ref{fig:lambda_redshift}). 

\section{Results}\label{sec:results}

We begin this Section by presenting the fitting performance of the five models used to fit the DM density profiles in Subsection~\ref{subsec:dm_profile_performance}, analysing the cumulative distribution function (CDF) of reduced the $\chi^2$. We then study in Subsection~\ref{subsec:cm_relation} the $c-M$ relation of NIHAO galaxies and quantitatively compare their distribution with the one obtained from SPARC observations using the KDE method described in \ref{subsubsec:comparison_sparc}. We furthermore investigate the correlation between $c_{200}$ and $M_{200}$ before ending this Subsection with an analysis of the effect of AGNs on the DM distribution (and thus $c_{200}$) of NIHAO galaxies.

We conclude this Section with the results on the stellar kinematics of the massive subset of NIHAO galaxies in Subsection~\ref{subsec:results_stellarkinematics}. We start by providing an example of how the features in the velocity maps of two galaxies are impacted by the presence of an AGN in their center. We then compute the distribution of NIHAO FRs and SRs in the angular momentum versus ellipticity plane and compare our results with \citet[][]{frigo_2019_mnras}, addressing the possible origins for the differences observed in ellipticities. The impact of AGN on the stellar kinematics of NIHAO galaxies is further determined by probing the radial evolution of their angular momentum at $z=0$, and our results are compared to observations from \citet[][]{atlas3d_iii}. Finally, we show the clear repercussion of AGN on the time evolution of the angular momentum of two example galaxies.

\subsection{DM profiles}\label{subsec:dm_profile_performance}

\begin{figure}
\centering
\includegraphics[width=\columnwidth]{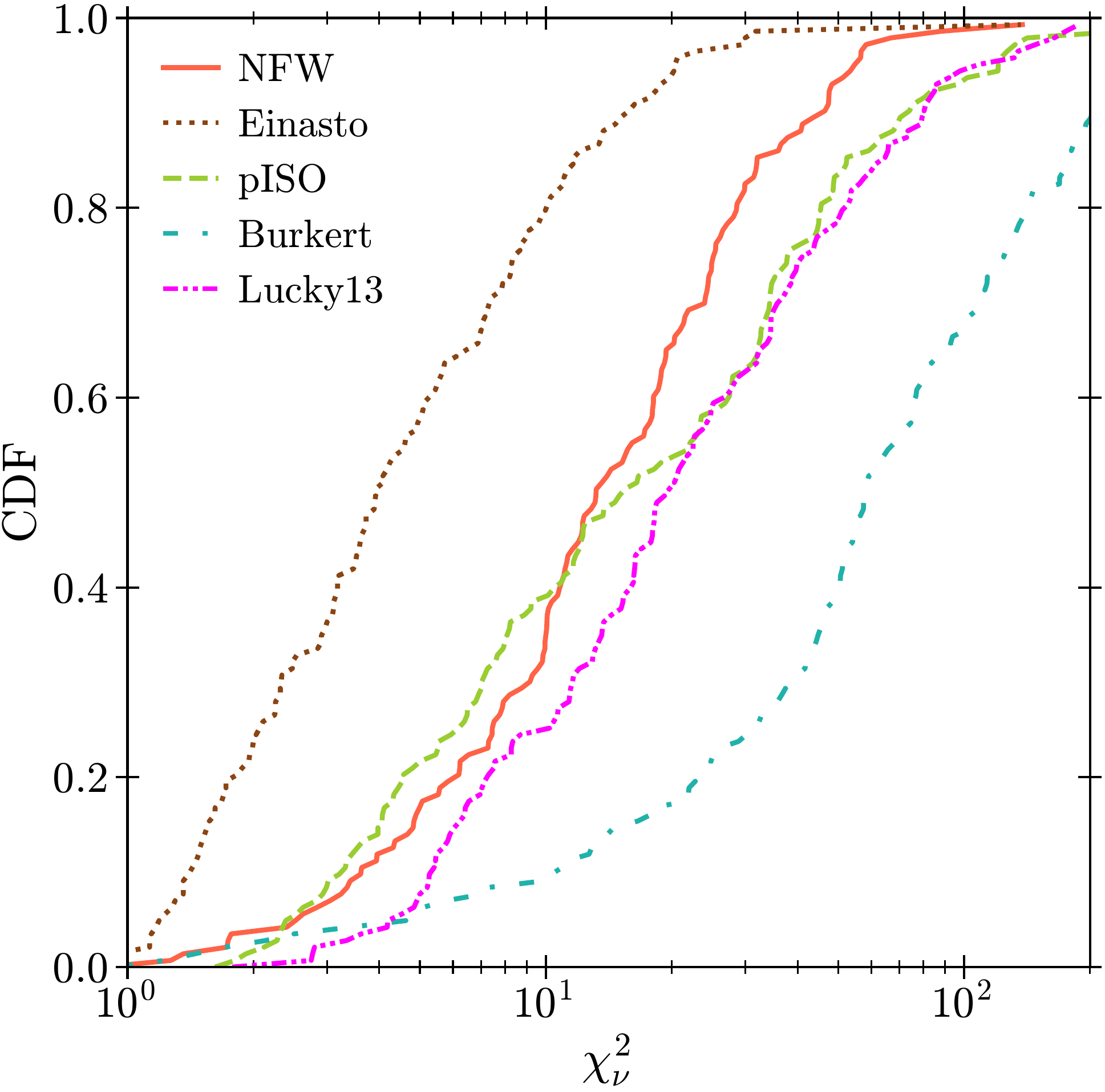}
\caption{Cumulative distribution function of the reduced chi-square $\chi_\nu^2$ of each profile. Each profile is shown in a different color: NFW (orange), Einasto (brown), pISO (light green), Burkert (light blue), and Lucky13 (magenta).}
\label{fig:cdf}
\end{figure}

In Fig.~\ref{fig:cdf}, the cumulative distribution function of the reduced chi-square $\chi^2_\nu = \frac{\chi^2}{N_{\text{b}}-p}$ is plotted for the five models, each identified by a different colour and line type. In the definition of $\chi^2_\nu$, $N_{\text{b}}$ designates the number of data points (i.e., bins), and $p$ is the number of parameters (two for NFW, pISO, Burkert, and Lucky13, and three for Einasto). Einasto performs the best, while Burkert is the worst. Our result for Einasto is consistent with the results of the CDF analysis of L20. Indeed, the additional free parameter allows the Einasto model to be more versatile with respect to the central core/cusp component of the DM profile.

After Einasto, in order of fitting performace, we have NFW followed by pISO, closely tailed by Lucky13. The success of NFW is due to the presence of several cuspy haloes accross the mass spectrum covered by NIHAO \citep[see for example figure 1 in ][]{Maccio2020}. The core radius in the pISO model can be set arbitrarily small, hence allowing it to show a versatile behaviour similar to Einasto. Lucky13 also seems to be capable of adapting to a majority of cuspy profiles, even though it was designed for central cores. Finally, Burkert is the model that performs the worst, which is to be expected since it cannot significantly vary its central slope and will inevitably fail at fitting cuspy profiles.

\subsection{$c-M$ relation}
\label{subsec:cm_relation}

In this Subsection, we are going to address three main aspects related to the $c-M$ relation in NIHAO and SPARC: (i) overall distribution of NIHAO galaxies in the $c-M$ plane, (ii) decreasing $c-M$ relation with halo mass, and (iii) impact of AGN on the concentration $c_{200}$.

\begin{figure*}
\centering
\includegraphics[width=\linewidth]{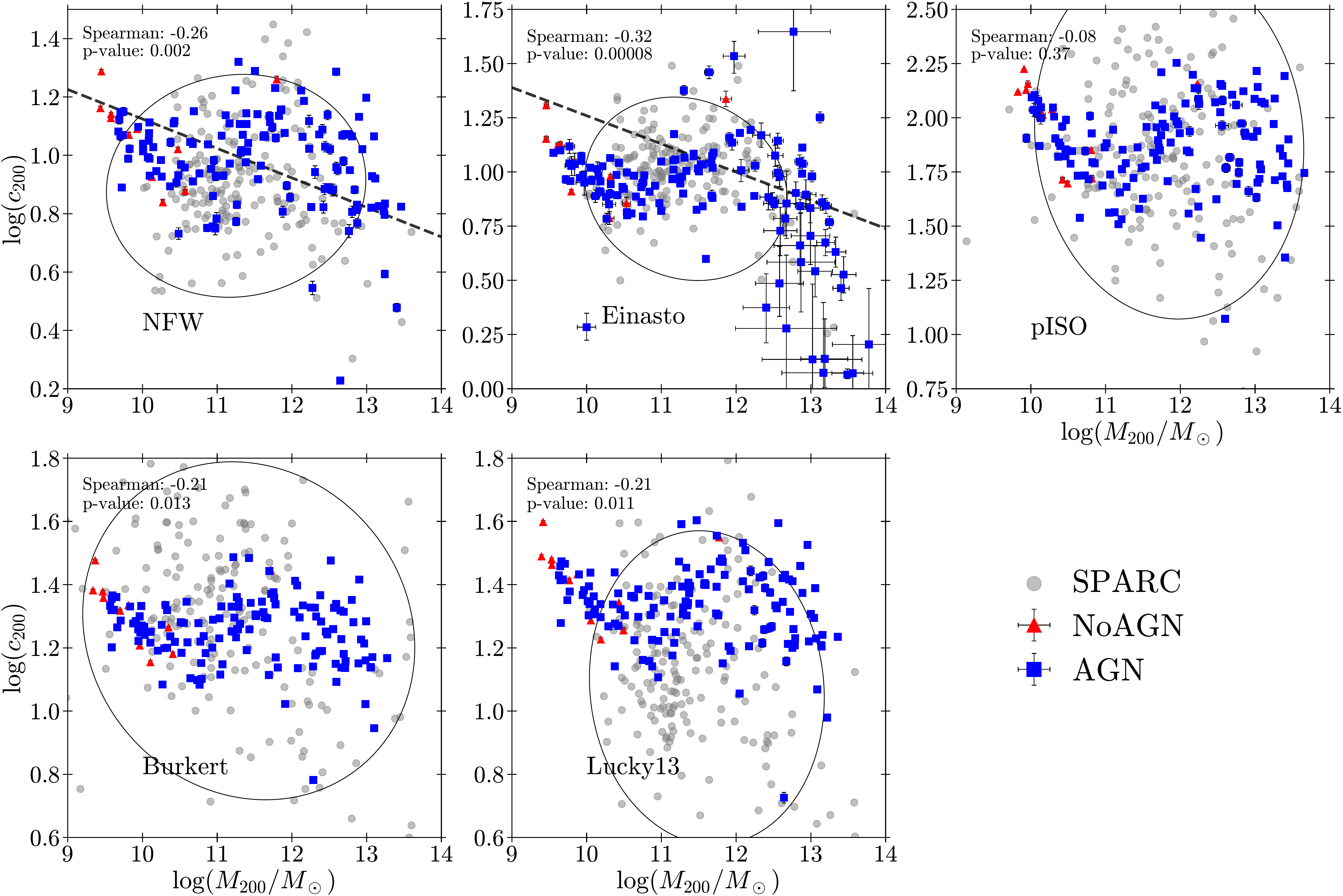}
\caption{$c-M$ obtained from our fits. The halo mass $M_{200}$ is plotted against the halo concentration $c_{200}$ for each of the five fitting models. NoAGN (red triangles) and AGN (blue squares) are shown alongside the data from L20 (grey dots) for visual comparison. For the 91 galaxies having both a NoAGN and AGN simulation, we use the AGN one. The NFW and Einasto panels additionally include the expected $c-M$ relation from \citet{maccio_2008_mnras}. The Spearman's correlation coefficients of NIHAO galaxies with their associated $p$-values are displayed for each profile in the top left corner. The black ellipses encompass the region chosen for the statistical comparison between NIHAO and SPARC following the method outlined in \ref{subsubsec:comparison_sparc}.}
\label{fig:m200c200}
\end{figure*}

\subsubsection{Distribution of NIHAO galaxies in the $c-M$ plane}

Given the free parameter $r_{\text{s}}$ and halo virial radius $R_{200}$ extracted from each profile, we compute the concentration of a halo as $c_{200} = \Rtwohun/r_{\text{s}}$ \citep[see, e.g.][for a review]{cooray_2002,okoli_2017}.
The concentration and mass $M_{200}$ obtained for the systems in NIHAO dataset are collected in Fig.~\ref{fig:m200c200}, together with the ones obtained for the SPARC dataset, as presented in L20. The red triangles represent the NoAGN galaxies, while the NIHAO galaxies with the presence of AGN are depicted in blue squares. In the background, the grey dots represents the results of the SPARC galaxies (Fig.~3 in L20). The dashed black line shows the expected $c-M$ relation from $N$-body simulations for the Einasto and NFW profiles \citep{maccio_2008_mnras}. The uncertainties on $M_{200}$ and $c_{200}$ are estimated following the method described in \ref{subsubsec:error} and are in most cases smaller than the points themselves. The most massive NIHAO AGN galaxies, however, show large uncertainties for Einasto arising from the MCMC fit. We have also added in each panel the fitted ellipse delimiting the KDE domain for the statistical comparison between NIHAO and SPARC.

\begin{figure*}
\centering
\includegraphics[width=\linewidth]{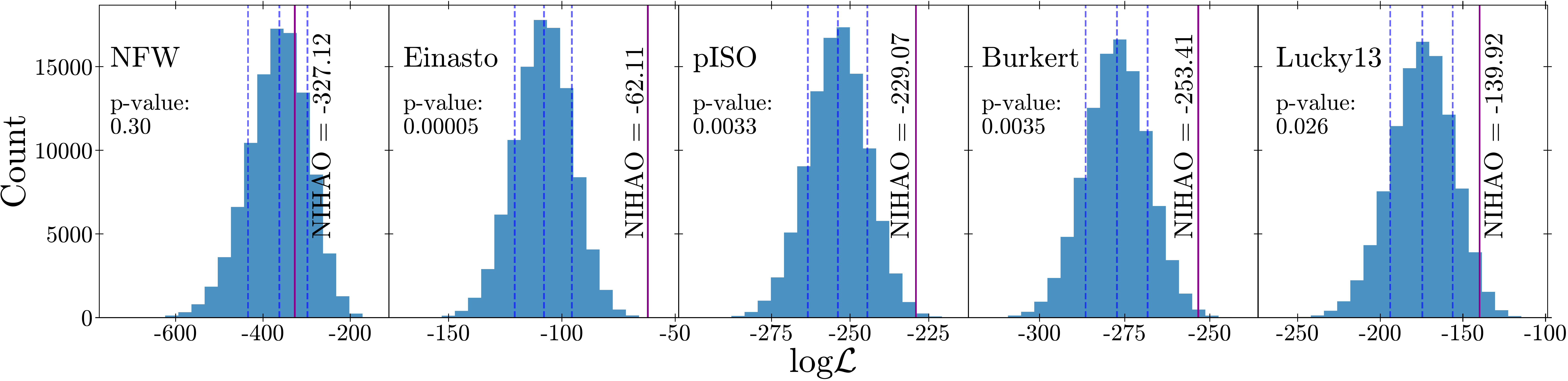}
\caption{Histograms of $\log{\mathcal{L}_k}$ for $k=10^{5}$ iterations of uniformly distributed random sets of points, grouped in 20 bins. Each panel represents one of the five profiles used, indicated in the upper left part. For reference, the blue dotted lines represent the 16th, 50th, and 84th percentiles of the score distribution. The magenta solid line shows $\log{\mathcal{L}_{\text{NIHAO}}}$ and we also added the $p$-value, i.e. the fraction of random draws scoring better than NIHAO.}
\label{fig:kde_hist}
\end{figure*}

Visually, the $c-M$ distribution extracted from the NIHAO sample is in good agreement with the 175 galaxies from SPARC. Quantitatively, we assess how close the two distributions are to each other using the KDE method outlined in \ref{subsubsec:comparison_sparc}, which confirms this agreement, as seen in Fig.~\ref{fig:kde_hist}.

A histogram of $\log{\mathcal{L}_k}$ for each of the profiles used in this work is plotted with blue dotted lines indicating the 16th, 50th, and 84th percentiles of the scores distribution. We computed the log likelihood of a uniformly distributed random set of points and repeated the procedure for $k=10^{5}$ iterations. The final score distribution is compared to $\log{\mathcal{L}_{\text{NIHAO}}}$, shown as a magenta solid line in each panel. The distance between the peak of the distribution and the NIHAO score in Fig. \ref{fig:kde_hist} assesses how different the log likelihood of NIHAO galaxies is compared to a uniform distribution of a set of random points. Except for NFW, $\log{\mathcal{L}_{\text{NIHAO}}}$ is systematically on the right of the histogram, indicating that the NIHAO $c-M$ relation is closer to SPARC than random points.

A more quantitative way to represent this result is obtained from the $p$-values, which amount to: 0.30, 0.000005, 0.0033, 0.0035, and 0.026 for NFW, Einasto, pISO, Burkert, and Lucky13, respectively. This number represents the fraction of the $10^5$ iterations that scored better than NIHAO. In other words, the probability that a uniform random distribution of points (within the domain defined by the ellipse) scores better than NIHAO against SPARC data is 30~per~cent for NFW, below 1~per~cent for Einasto, pISO, and Burkert respectively, and below 5~per~cent for Lucky13. While crude, this approach rejects the null hypothesis that NIHAO is no different from random with high confidence for most halo models.

\subsubsection{Correlation between $c_{200}$ and $M_{200}$}

We move on to the next interrogation asked in the beginning of this subsection, namely does the concentration fall with increased halo mass? As previously mentioned, \citet{navarro_1996_apj} found using $N$-body simulations that the concentration of a DM halo is related to its mass, with higher mass haloes exhibiting a lower concentration than their less massive counterparts. This inverse correlation arises from the dependence between the central density of a collapsed halo and the initial density distribution of the same region at the epoch of collapse \citep[][]{navarro_1997_apj,zhao_2003_mnras,zhao_2003_apj}. Since small-scale structures collapse earlier, their concentration is expected to be higher, reflecting the higher background density of the central region when it collapsed \citep[][]{wechsler_2002_apj}

Similarly to L20, we tackle this question by computing Spearman's correlation coefficients for each model as well as their associated $p$-value, which measure the rank correlation between two variables and their corresponding $p$-value under the null hypothesis that the galaxies are uncorrelated in the $c-M$ plane. Showcased in the top left corner of each panel in Fig.~\ref{fig:m200c200}, our correlation coefficients are -0.26, -0.32, -0.08, -0.21, and -0.21 with $p$-values 0.002, 0.00008, 0.37, 0.013, and 0.011 for NFW, Einasto, pISO, Burkert, and Lucky13, respectively. pISO is the only profile where the null hypothesis cannot be rejected and is thus consistent with no correlation in the NIHAO data. The four remaining profiles show significant but weak anti-correlations below the five~per~cent level (Burkert and Lucky13) and one~per~cent one (NFW and Einasto). Therefore, despite our data showing significant anti-correlation in four out of five profiles, the large scatter observed in concentration (and thus weak Spearman's coefficients) does not allow for a definitive answer to the question.


\subsubsection{Impact of AGN on $c_{200}$}
\begin{figure}
\centering
\includegraphics[width=\columnwidth]{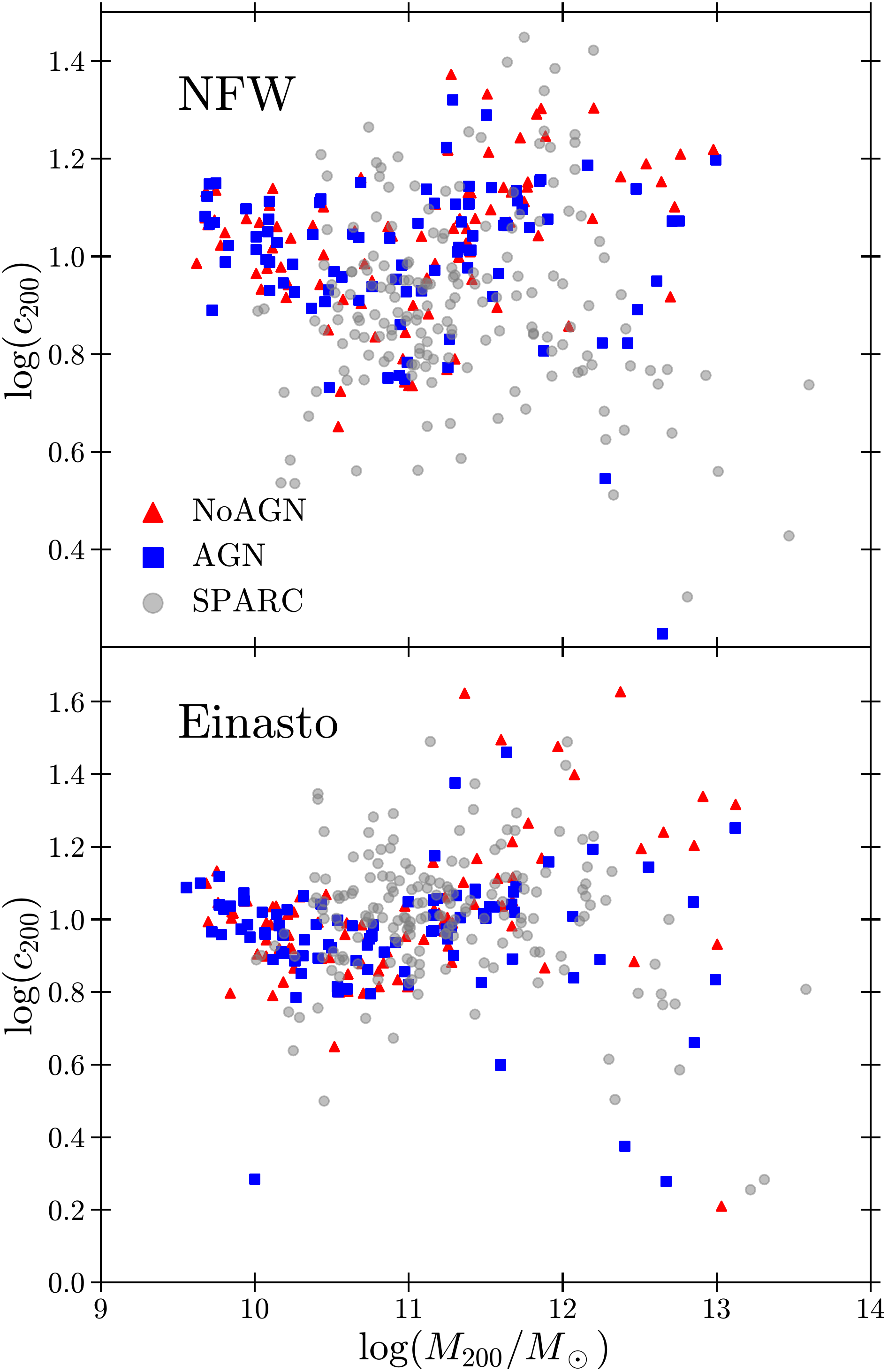}
\caption{$c-M$ obtained from our fits. The halo mass $M_{200}$ of the 91 NIHAO simulated both with and without AGN is plotted against the halo concentration $c_{200}$ for the NFW and Einasto models. NoAGN (red triangles) and AGN (blue squares) are shown alongside the data from L20 (grey dots) for visual comparison.}
\label{fig:m200c200_2}
\end{figure}

Finally, we use the 91 NIHAO simulations run both with and without AGN to investigate if the AGN itself brings concentration down compared to galaxies devoid of a central BH and the associated physical processes. A first glimpse of the effect of AGN on the concentration is displayed in Fig.~\ref{fig:m200c200_2}. Similarly to Fig.~\ref{fig:m200c200}, we plot the $c-M$ relation for NIHAO NoAGN as red triangles and their AGN counterpart as blue squares for both NFW and Einasto profiles. SPARC galaxies are shown again as grey dots for visual comparison. Below $\log(M_{200}/ \text{M}_{\odot}) \sim 11.5$, both NoAGN and AGN samples show a similar distribution indicating that AGN have no significant effect on low-mass galaxies. Above $\log(M_{200}/\text{M}_{\odot}) \sim 11.5$, however, the degeneracy no longer holds as both distributions begin to depart from each other and the AGN simulations trend towards lower concentrations.

\begin{figure}
\centering
\includegraphics[width=\columnwidth]{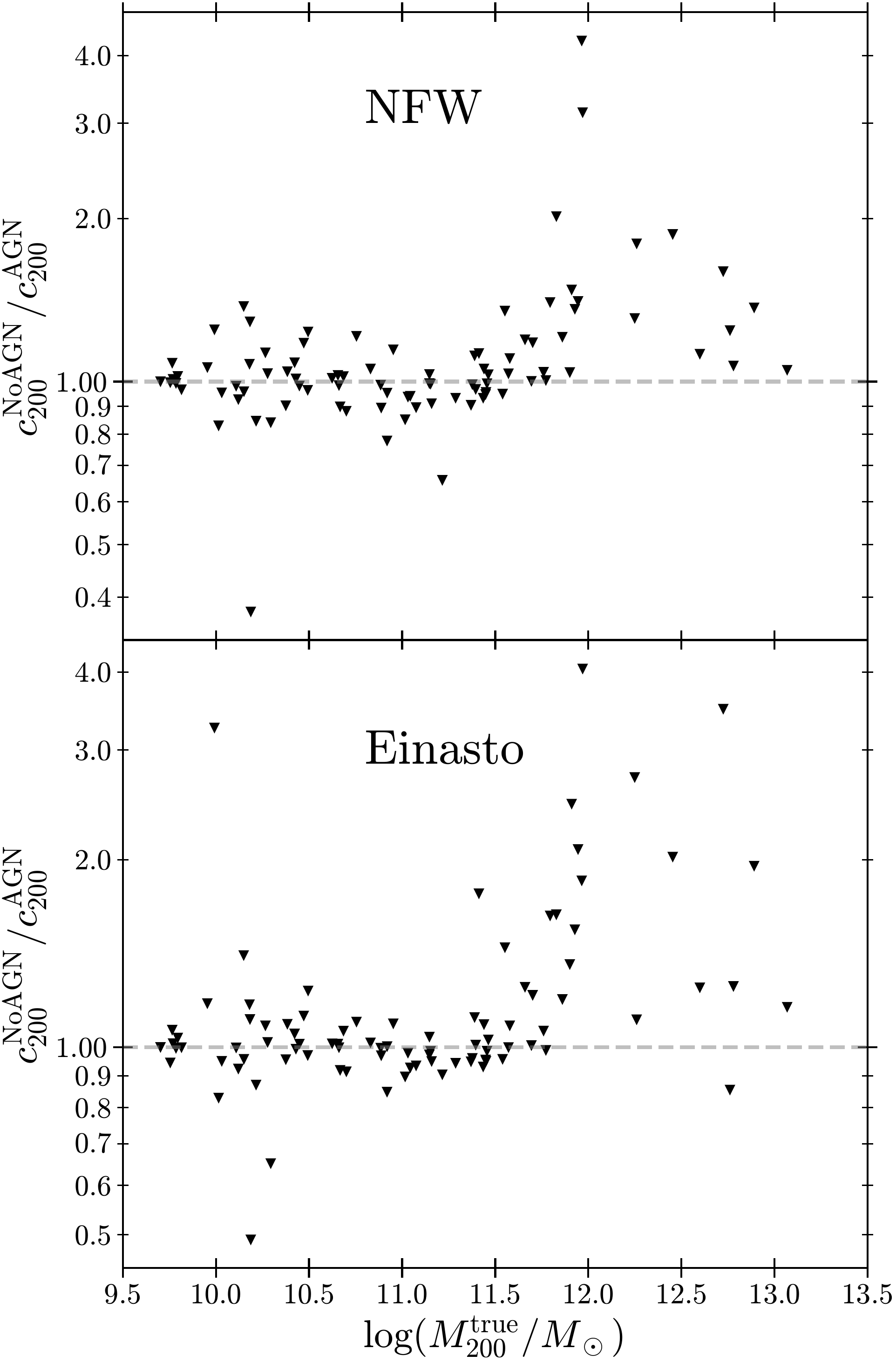}
\caption{NIHAO NoAGN concentration divided by AGN concentration as a function of NoAGN halo mass for the 91 pairs of simulations run both with and without AGN. The top panel shows the result for NFW and the bottom panel shows the result for Einasto. A grey dotted line at $y=1$ is added for clarity: all the points above the line indicate that the inclusion of AGN physics resulted in a lower concentration for the same halo.}
\label{fig:m200c200_3}
\end{figure}

These results are furthermore confirmed in Fig.~\ref{fig:m200c200_3}. For each pair of simulations, we select the `true' halo mass $M_{200}^{\text{true}}$ of the AGN simulation as the reference mass plotted in the $x$-axis, and we plot the corresponding ratio between the NoAGN concentration and the AGN one, showed as black triangles for NFW (top panel) and Einasto (bottom panel). This representation highlights the effect of the AGN presence on concentration. For clarity, we also add a grey dotted line at $y=1$: all points above this line indicate that the AGN simulation has a lower concentration than the NoAGN one.

Fig. \ref{fig:m200c200_3} shows that for NFW, 38 out 91 simulations are below the dotted line. The Einasto profile displays similar results with 37 out of 91 simulations resulting in a mostly slight increase in concentration when AGN is added. Splitting again our analysis between galaxies located below and above the $10^{11.5} \text{M}_{\odot}$ halo mass threshold confirms what is observed in Fig.~\ref{fig:m200c200_2}. For both NFW and Einasto, our simulations are distributed relatively uniformly around the dotted line before showing a clear upward trend for high mass galaxies. More specifically, only 1 out of 27 and 3 out of 27 galaxies are found below the $y=1$ line for NFW and Einasto, respectively. These results mean that high-mass galaxies AGNs lower the concentration in almost all cases, with the concentration ratio of the majority of haloes lying between 1 and 2, and the most extreme cases showing a reduction in concentration by a factor of 4.

\begin{figure}
\centering
\includegraphics[width=\columnwidth]{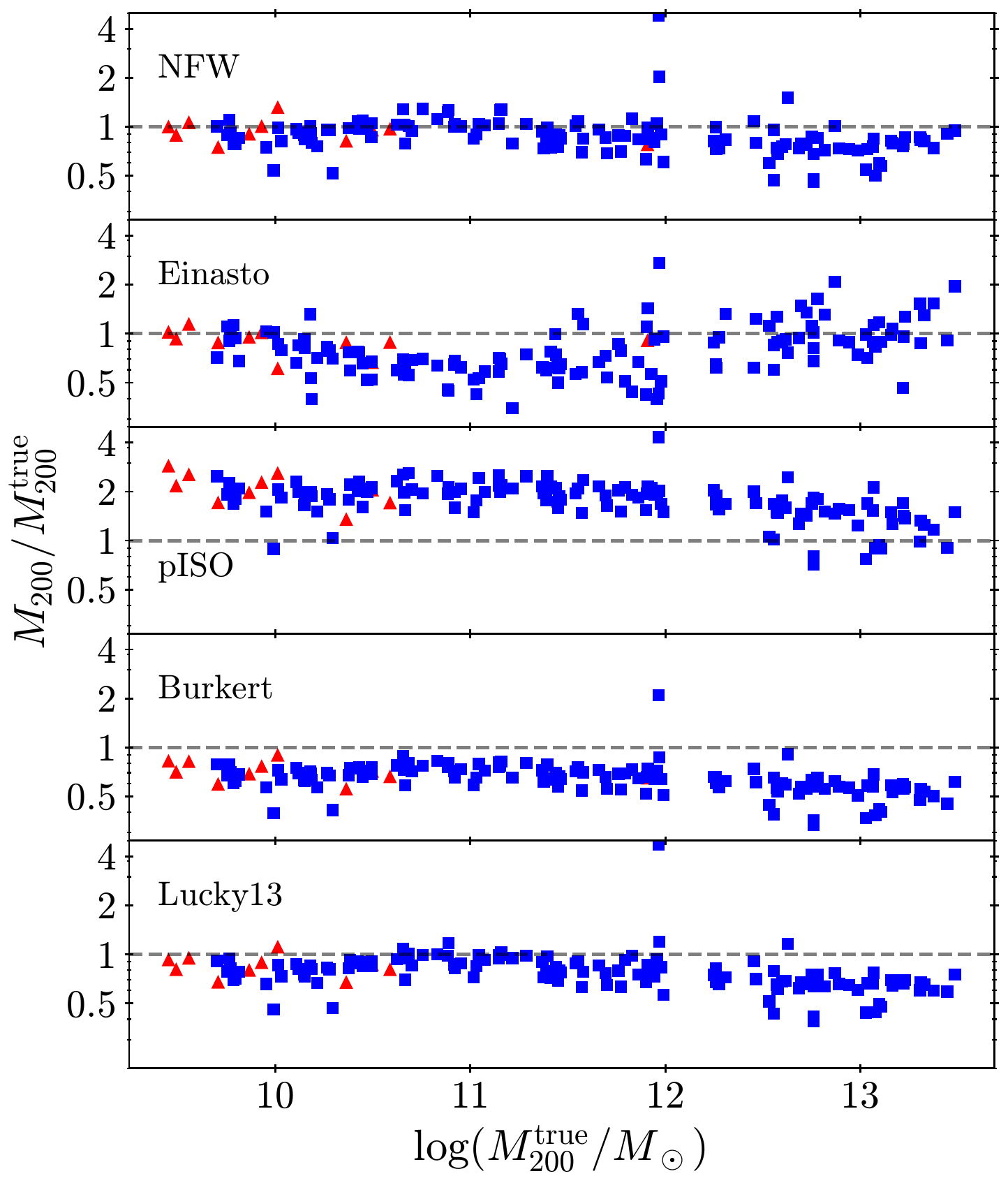}
\caption{Ratio between $M_{200}$ obtained from the fit of each profile and the `true' mass from the simulation as a function of the simulation halo mass. Galaxies without AGN are represented as red triangles, while AGN ones are displayed as blue squares. The grey dotted line indicates where the ratio is 1.}
\label{fig:m200_m200_fit}
\end{figure}

To conclude the study on the dark components, we check for the fact that the halo masses we used in the $c-M$ relation are not biased by performing a comparison between the halo mass obtained from fitting the different DM profiles ($M_{200}$ used in the $c-M$ relation) and the `true' halo mass from the simulations themselves ($M_{200}^{\text{true}}$). The results are showcased in Fig.~\ref{fig:m200_m200_fit}. Each panel corresponds to one DM profile and the ratio $M_{200}/M_{200}^{\text{true}}$ is plotted against $M_{200}^{\text{true}}$. NoAGN galaxies are represented as red triangles, while AGN ones are displayed as blue squares. A grey dotted line is added for clarity where both masses are equal. Apart from a few outliers, our calculated masses are consistent with $M_{200}^{\text{true}}$. Each profile has a tendency to slightly undershoot the halo mass however, except for pISO which shows constant overshooting by a factor of 2.

The difference observed from one profile to another can be explained as follows. As $M_{200}$ is an integrated quantity, the outer bins of the DM density profile contain a large fraction of the total mass. A slight difference in $r_{\text{s}}$ obtained from fitting different profiles can propagate and lead to a larger divergence of the profiles around $R_{200}$, thus inducing distinct results for $M_{200}$. In addition, we compute our profiles up to only 20~per~cent of the `true' virial radius, as mentioned in \ref{subsubsec:virialrad_virialmass}. With almost all NIHAO galaxies located in the 0.5-2 $M_{200}/M_{200}^{\text{true}}$ range in Fig.~\ref{fig:m200_m200_fit}, we are confident that our estimated masses give a fair representation of the `true' simulated ones.

\subsection{Stellar kinematics}\label{subsec:results_stellarkinematics}
\begin{figure*}
\centering
\includegraphics[width=\linewidth]{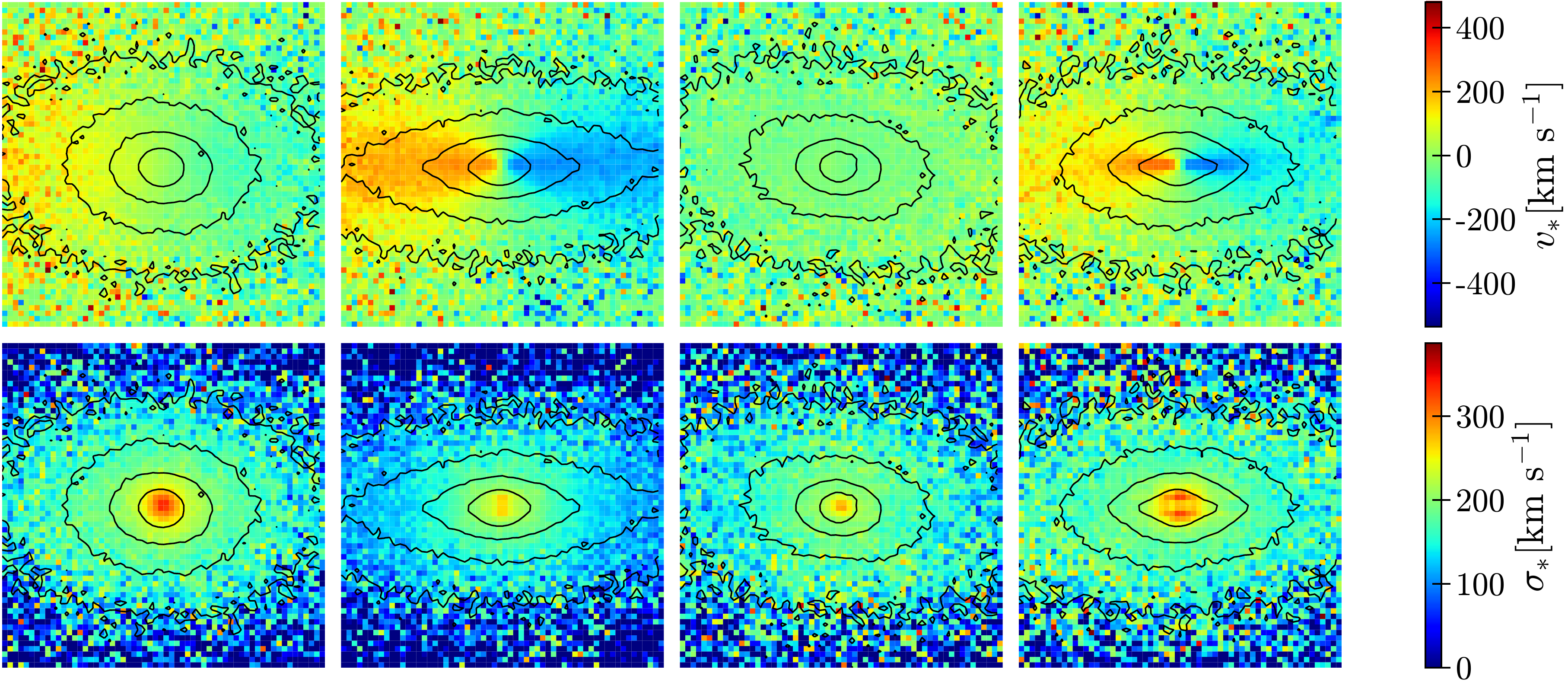}
\caption{Stellar velocity (upper panels) and velocity dispersion (lower panels) maps in edge-on view for two NIHAO massive elliptical galaxy (g5.41e12), along with isophotes. From left to right: g5.41e12 AGN, g5.41e12 NoAGN, g7.50e12 AGN, and g7.50e12 NoAGN. The side of each map is five times the projected stellar half-mass radius.}
\label{fig:velmaps}
\end{figure*}

We leave the dark sector and move to the visible motions of stars in this remaining subsection. Panels in Fig. \ref{fig:velmaps} show an example of the stellar line-of-sight velocity (upper) and velocity dispersion (lower) in edge-on view for two galaxies as an example (from left to right: g5.41e12 AGN/NoAGN and g7.50e12 AGN/NoAGN). The isophotes are displayed as black closed curves. The NoAGN cases show typical FR features with aligned photometry and kinematic axis. The effect of AGN is clear with both cases exhibiting rounder isophotes, less disky structure and suppressed rotation, similarly to what is exhibited in \citet[][]{frigo_2019_mnras} stellar kinematics maps. The aforementioned visible consequences are indicative of AGN feedback turning FRs into SRs.

\begin{figure}
\centering
\includegraphics[width=\linewidth]{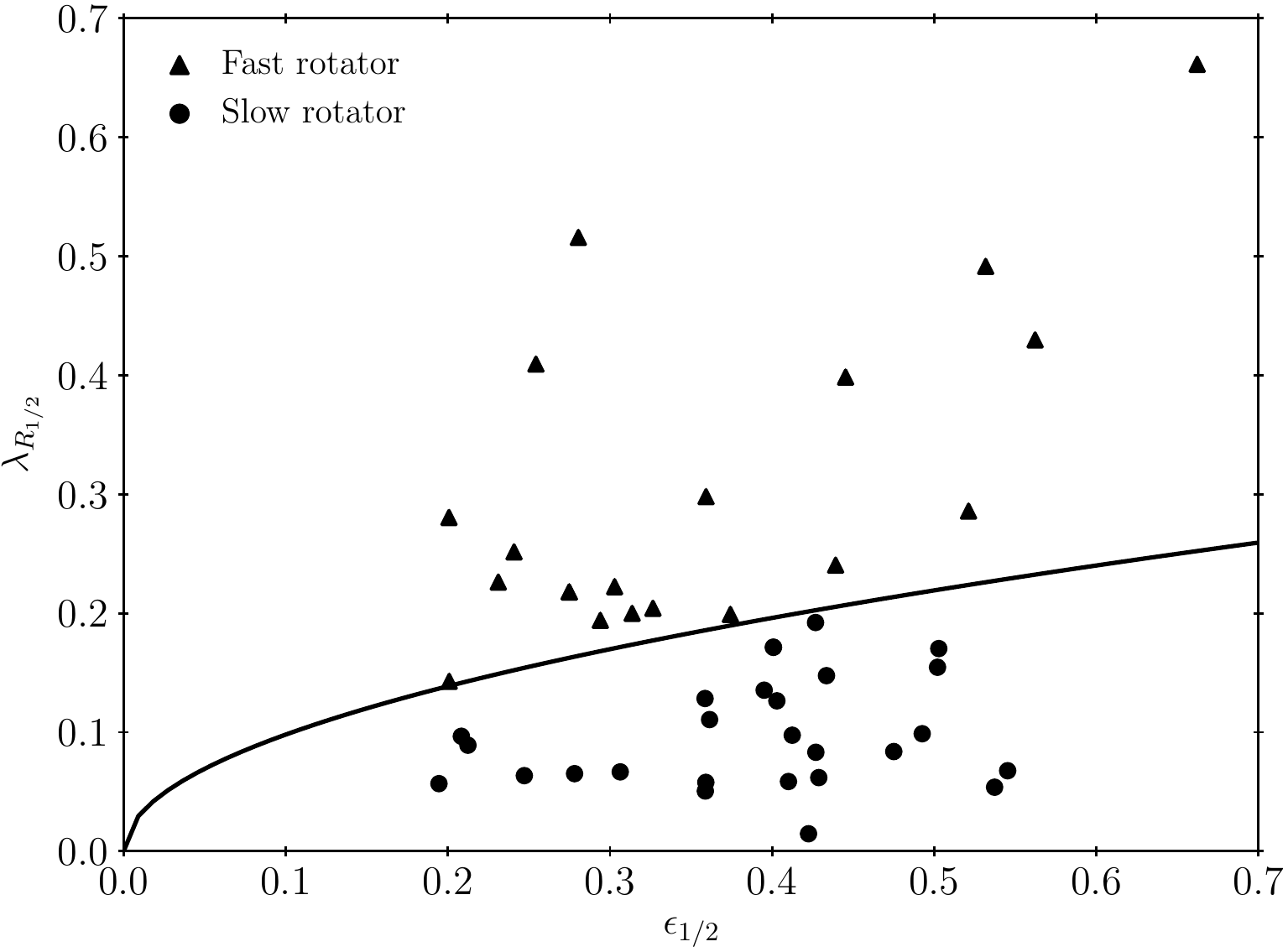}
\caption{Distribution of the 45 NIHAO massive AGN galaxies at $z = 0$ in the $\lambda_{R_{1/2}}-\epsilon_{1/2}$ plane. FRs are represented as black triangle while SRs as black circles. The black line defines the ATLAS$^{\text{3D}}$ cutoff between SRs and FRs.} 
\label{fig:lambda_epsilon}
\end{figure}

For our sample of 45 elliptical galaxies at $z = 0$, we begin by classifying each of them as either FRs or SRs. Galaxies lying below the cutoff line in the $\lambda_{R_{1/2}}-\epsilon_{1/2}$ plane are classified as SRs ($\lambda_{R_{1/2}} < 0.31 \sqrt{\epsilon_{1/2}}$), while the ones above as FRs ($\lambda_{R_{1/2}} > 0.31 \sqrt{\epsilon_{1/2}}$), as represented visually in Fig.~\ref{fig:lambda_epsilon}. FRs (black triangles) and SRs (black dots) are plotted in the $\lambda_{R_{1/2}}-\epsilon_{1/2}$ plane with the black curve determining the cutoff between SRs and FRs. We further verified that the observed abundances of FRs and SRs do not considerably change by varying the viewing angle, as expected given our use of equation~(\ref{eq:lambda_crit}) to kinematically classify the galaxies in our sample. The majority of our massive galaxies are slow rotating, in agreement with results from \citet[][]{frigo_2019_mnras}. Except for one galaxy, all  members of our sample have ellipticities below 0.6, which is slightly higher than the 0.4 limit obtained by \citet[][]{frigo_2019_mnras}. On the other hand, given that our work used a different code, feedback scheme, and a partially different way to compute $\epsilon_{1/2}$, we do expect some departures in the parameters of the two galaxy populations. The distribution of the NIHAO galaxies in the $\lambda_{R_{1/2}}-\epsilon_{1/2}$ plane is continuous, in agreement with observational results \citep{atlas3d_iii} and we notice that on average, our SRs have higher ellipticities than observed, a trend also present in previous simulations \citep[e.g.][]{atlas3d_xxv}.


\begin{figure}
\centering
\includegraphics[width=\linewidth]{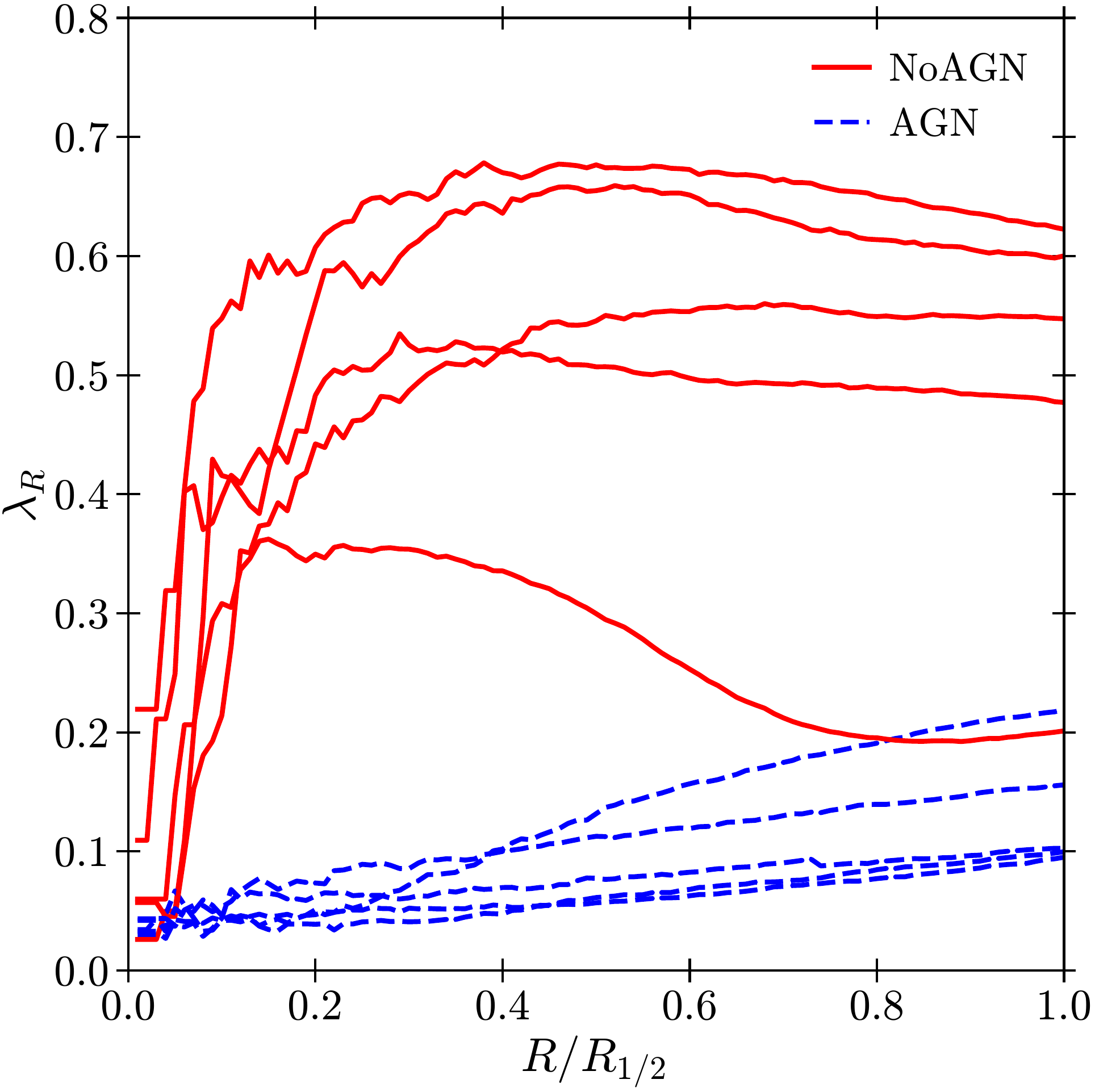}
\caption{Spin parameter profile up to one projected stellar half-mass isophote for five massive galaxies with (blue dashed lines) and without (red continuous lines) AGN feedback in edge-on view at $z=0$.}
\label{fig:lambda_profile_radius}
\end{figure}

In order to investigate the impact of AGN on the stellar kinematics of NIHAO massive galaxies, we compute $\lambda_R$ in edge-on view as a function of radius relative to one projected stellar half-mass radius for five of our massive elliptical galaxies at $z=0$, as shown in Fig. \ref{fig:lambda_profile_radius}. The red curves represent galaxies without AGN whereas the blue curves are for the same galaxies with AGN physics. The galaxies' masses are of the order $10^{12}$ - $10^{13}$ \Msun. The presence of AGN in the simulated galaxies has a strong impact on the spin parameter, maintaining it at relatively low values. All AGN galaxies exhibit a slow and steady increase of their angular momentum up to about $\lambda_R \sim 0.1-0.2$ at the projected stellar half-mass radius. Introducing AGN physics leads to significantly lower values of $\lambda_R$ for all simulations, turning all five galaxies into SRs. In contrast, only one out of five galaxies is classified as a SR when AGN feedback is removed. 

When compared with results from ATLAS$^{\text{3D}}$ sample \citep[][]{atlas3d_iii}, the role of AGN feedback becomes even more important: without AGN  less than 10~per~cent of our galaxies are classified as SRs and when AGN is included, this fraction grows above 57~per~cent (26/45). Despite obtaining a higher fraction of SRs than the observed 34~per~cent, it is clear from our simulations that SRs do not readily form in a simulated universe without AGN feedback, in accordance with previous findings \citep[e.g.][]{atlas3d_xxv,wu_2014_mnras,frigo_2019_mnras}.

\begin{figure}
\centering
\includegraphics[width=\linewidth]{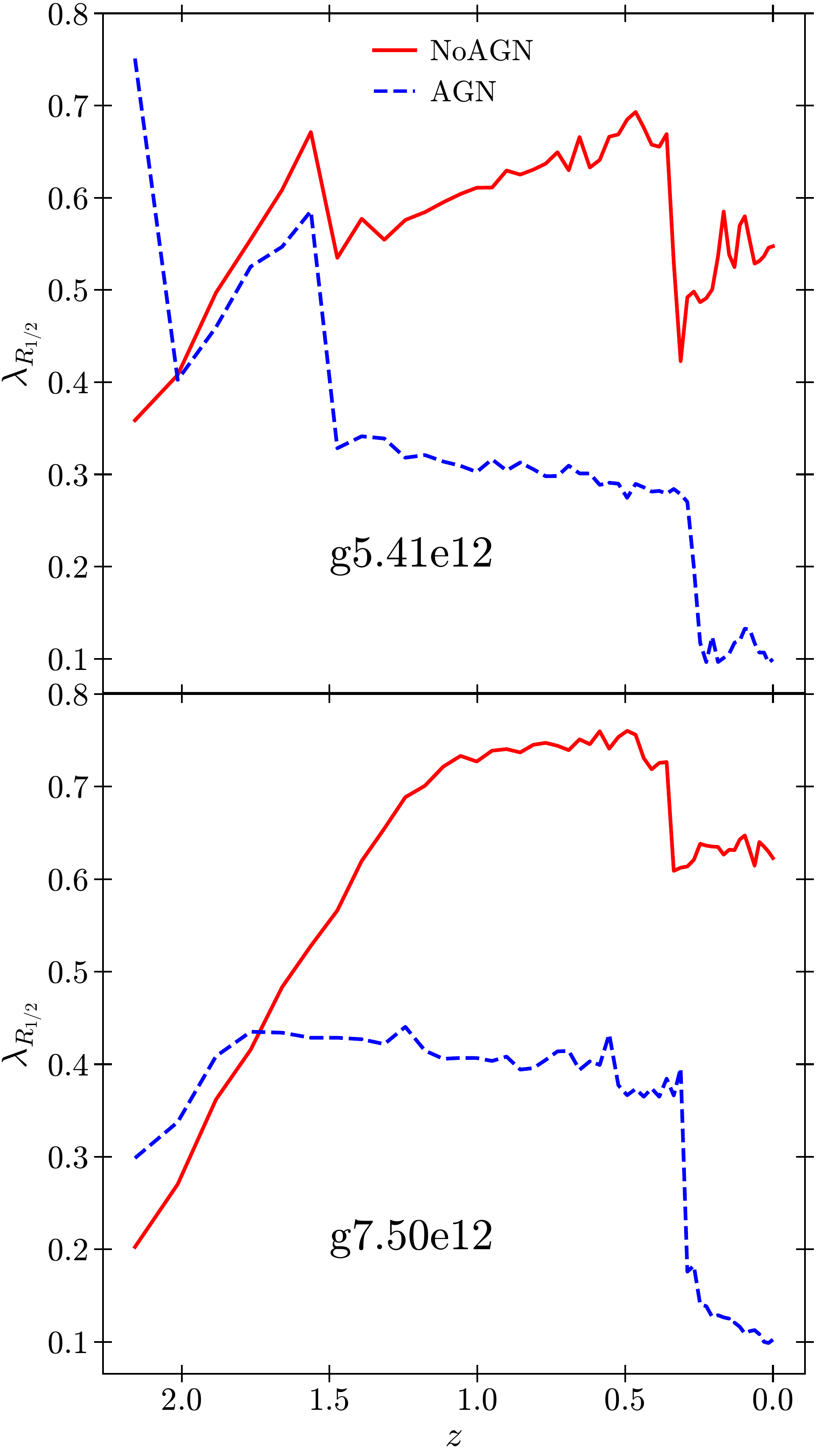}
\caption{The evolution of the spin parameter for two massive NIHAO galaxies with (blue dashed lines) and without (red continuous lines) AGN feedback, in edge-on view as function of redshift $z$. The names of the simulations are indicated in the lower central part of both panels.}
\label{fig:lambda_redshift}
\end{figure}

In order to better understand the effects of AGN, we
trace the evolution of the stellar kinematics through cosmic time for a subset of two massive elliptical galaxies with and without AGN physics. We compute the redshift evolution of $\lambda_{R_{1/2}}$ from $z \sim 2.1$ to $z=0$ for each pair of galaxies in edge-on view, as shown in Fig. \ref{fig:lambda_redshift}. The trend that AGN feedback results in a reduction of the galaxy spin is once again evident.

On the top panel, both NoAGN and AGN are FRs around $z=1.5$ with $\lambda_{R_{1/2}}$ slightly below 0.7 and 0.6, respectively. A merger event leads to a drop of the spin parameter, with the NoAGN case recovering some rotation until a second merger at $z \sim 0.3$, when $\lambda_{R_{1/2}}$ is able to increase again after the merger. On the contrary, $\lambda_{R_{1/2}}$ in the AGN case continues to fall after the first merger and the galaxy ends up rotating very slowly with $\lambda_{R_{1/2}} \sim 0.1$. The second example in the bottom panel shows a similar behaviour where the spin parameter steadily increases in both case, plateauing around 0.7 (NoAGN) and 0.4 (AGN). After a merger at $z \sim 0.3$, rotation is slightly slowed for the galaxy without AGN, reaching $\lambda_{R_{1/2}} \sim 0.6$, while the AGN case encounters a significant drop and keeps falling afterward to reach $\lambda_{R_{1/2}} \sim 0.1$.

\section{Discussion and conclusion}\label{sec:discussion}

In this paper, we investigated impact of AGN on DM profiles and stellar kinematics of simulated haloes in the NIHAO \citep[][]{nihao1} simulation suite. We recovered a distribution of galaxies in the $c-M$ plane, which is consistent with the one derived from observations of the SPARC survey. However, NIHAO galaxies tend to show less scatter in concentration, especially for the pISO, Burkert, and Lucky13 profiles. The tighter $c-M$ relation in NIHAO is probably caused by the observational and fitting errors arising from indirectly inferring the underlying DM distribution from rotation curves, which is instead readily available in simulations. In spite of this difference, we reach analogous conclusions as L20 when calculating the Spearman's correlation coefficients of each model. Both simulations and observations do not show a strong (anti)correlation between the concentration parameter and the halo mass, and hence both approaches do not reproduce in details the expected
relation predicted by  $N$-body simulations 
\citep[i.e.][]{maccio_2008_mnras}, showing the importance of  baryons in altering the expectations from a pure CDM universe.



Fig.~\ref{fig:m200c200} showcases an interesting u-shaped feature in the $c-M$ relation centered around $M_{200} \sim 10^{10}-10^{11}\,\Msun$, visible in all five panels. Since below $M_{200} \sim 10^{12}\,\Msun$ AGNs have no impact on the concentration (see Figs.~\ref{fig:m200c200_2} and \ref{fig:m200c200_3}), the origin of this characteristic is found in baryonic interactions with DM through supernova feedback and was previously studied from the NIHAO catalogue by \citet[][]{Tollet2016}. Starting from $M_{200}$ between $10^9$ and $10^{10}$, the concentration decreases as the halo grows in mass and thus also in stellar mass, making supernova feedback more efficient at heating gas and expanding it. The expansion of gas can induce a change in the underlying gravitational potential which, if rapid enough, can be responsible for a non-adiabatic expansion of DM particles \citep[see, e.g.][]{pontzen_2012_mnras}. Above $M_{200} \sim 10^{11}\,\Msun$, haloes become too massive and supernova feedback itself cannot provide enough energy, and concentration goes back up again.


Above $M_{200} \sim 10^{12}\,\Msun$, AGN feedback takes over and the concentration starts falling again \citep[see, e.g.][]{nihao23}. We probed how significant of an impact do AGNs have in lowering the halo concentration with help of NIHAO simulations available in pairs, both with and without AGN. The clear effect of adding BH physics in the hydrodynamical simulations on the concentration is visible in Fig. \ref{fig:m200c200_2}: AGNs lower the concentration of their host halo by up to a factor of four. The physical explanation for this phenomenon is similar to that of supernova feedback. The energy outflow from the AGN feedback increases the temperature of the surrounding gas which is transported away from the central region through convection, thus varying locally the gravitational potential \citep[][]{martizzi_2012_mnras}.


On the more luminous side, we found good agreement in the distribution of SRs and FRs among massive galaxies with AGN compared to observations. Using the cutoff proposed by ATLAS$^{\text{3D}}$, we find almost 60~per~cent of massive NIHAO galaxies to be SRs, as shown in Fig. \ref{fig:lambda_epsilon}. When we compare the AGN and NoAGN simulations, the effect of AGN is clear: AGN feedback is required to reproduce the observed dichotomy between FRs an SRs in the Universe.


From Fig. \ref{fig:lambda_redshift}, we can observe that the occurrence of major mergers at $z \sim 1.5$ and $z \sim 0.3$ reduces the stellar angular momentum for both galaxies with and without AGN feedback. Following the event of major mergers, massive galaxies with AGN feedback are unable to accrete cold gas which, suppresses in-situ star formation \citep[][]{dubois_2013_mnras,martizzi_2014_mnras,penoyre_2017_mnras,choi_2018_apj,frigo_2019_mnras}. Therefore, galaxies with AGN are unable to form a new fast-rotating stellar component, which leads them to evolve into SRs. On the other hand, massive galaxies without AGN can regain some of their spin and remain FRs due to cold gas accretion and subsequent formation of a fast-rotating stellar disc. Thus, AGN feedback increases the abundances of SRs through the action of mergers, which also suggests that AGN feedback holds key to reproducing the observed abundances of FRs and SRs in massive galaxies.

To summarise, we demonstrated that the inclusion of AGN feedback in cosmological hydrodynamical simulations of galaxies leads to a decrease in $c$ of the DM halo. This behaviour was studied across five different fitting models, and the reduction in $c$ due to AGN is very clear when comparing high-mass galaxies with and without AGN. The absence of AGN in these galaxies leads to a higher $c$ for almost all of them for the NFW and Einasto profiles, as shown in Fig. \ref{fig:m200c200_3}. We were thus able to reproduce the $c-M$ relation to good agreement with observational data of 175 late-type galaxies from the SPARC database with similar scatter. We also confirmed previous results that Einasto stands as the preferred profile for accurately fitting halo DM density profiles.

Finally, following \citet{atlas3d_iii}, we are able to classify early-type galaxies into SRs and FRs using the threshold value for the spin parameter defined in Section \ref{sec:methods}. We show that AGNs have a direct impact on the stellar kinematics of these galaxies, preventing further gas accretion and suppressing in-situ star formation, thereby turning the host galaxies into SRs.

The addition of AGN feedback in comsological simulations has proven successful at retrieving the SFR of observed galaxies by quenching star formation that would otherwise produce simulated galaxies with an overabundance of stars. Here, we show that AGN feedback generates additional `collateral' effects in the underlying DM distribution of host haloes, as well as in the stellar kinematics of galaxies. These effects can be used to help further constrain the modelling of this type of feedback in simulations.

\newpage 

\section*{Acknowledgements}
We are grateful to the authors of L20 who kindly agreed to share their data with us. 
This material is based upon work supported by Tamkeen under the NYU Abu Dhabi Research Institute grant CAP$^3$.
The authors  gratefully acknowledge the Gauss Centre for Supercomputing e.V. (www.gauss-centre.eu) 
for funding this project by providing computing time on the GCS Supercomputer SuperMUC at Leibniz
Supercomputing Centre (www.lrz.de) and the High Performance Computing resources at New York University Abu Dhabi.
M.~P. acknowledges financial support from the European Union’s Horizon 2020 research and innovation program under the Marie Sklodowska-Curie grant agreement No. $896248$.

\section*{Data availability}
The data underlying this article will be shared on reasonable
request to the corresponding author.



\bibliographystyle{mnras}
\bibliography{ref} 





\bsp	
\label{lastpage}
\end{document}